\documentclass[12pt,preprint]{aastex}

\def\etal {\hbox{et al.}}
\def\Msol {\hbox{M$_{\odot}$}}

\def\mss {\hbox{mag arcsec$^{-2}$}}
\def\kms {\hbox{${\rm km\, s}^{-1}$}}
\def\acs {\hbox{$^{\prime\prime}$}}

\def\Ha {\hbox{H{$\alpha$}}}
\slugcomment{Submitted to The Astrophysical Journal}

\shortauthors{O'Neil et al.}
\shorttitle{Star Formation in LSB Galaxies }

\begin{document}
\title{Star Formation in Galaxies with Large Lower Surface Brightness Disks}
\author{K. O'Neil}
\affil{NRAO, PO Box 2, Green Bank, WV 24944}
\email{koneil@nrao.edu}

\author{M. S. Oey}
\affil{University of Michigan,  Astronomy Department, 830 Dennison Building Ann Arbor, MI 48109-1042}
\email{msoey@umich.edu}

\and

\author{G. Bothun}
\affil{University of Oregon, Physics Department, 1371 E 13th Avenue, Eugene, OR 97403}
\email{nuts@bigmoo.uoregon.edu}

\begin{abstract}
We present B, R, and \Ha\  imaging data of  19 large disk galaxies
whose properties are intermediate between classical low surface
brightness galaxies and ordinary high surface brightness galaxies.
We use data taken from the Lowell 1.8m Perkins telescope to determine the
galaxies' overall morphology, color, and star formation properties.
Morphologically, the galaxies range
from Sb through Irr and include galaxies with and without nuclear bars.
The colors of the galaxies vary from B$-$R = 0.3 -- 1.9, and most show at least
a slight bluing of the colors with increasing radius.
The \Ha\ images of these galaxies show an average star formation rate lower
than is found for similar samples with higher surface brightness disks.  Additionally,
the galaxies studied have both higher gas mass-to-luminosity and
diffuse \Ha\ emission than is found in higher surface brightness
samples. 
\end{abstract}

\section{Introduction}

Large low surface brightness galaxies are galaxies
with disk central surface brightnesses statistically far from the 
\citet{freeman70} value of $\mu_B(0)$ = 21.65 $\pm$ 0.3 \mss, and 
whose properties are significantly removed from the dwarf galaxy category
(e.g. $M_B < -18$, $M_{\rm HI} > 10^9$\Msol).  Studies of large LSB
galaxies have discovered a number of intriguing facts:
large LSB galaxies, in contrast to dwarf LSB galaxies, can exhibit
molecular gas \citep{das06, oneil04b, oneil03, oneil00b};
the gas mass-to-luminosity ratios of large LSB galaxies are typically 
higher than for similar high surface brightness counterparts by a factor of 
2 or more \citep{oneil04}; and, like dwarf LSB galaxies, large LSB systems 
are typically dark-matter dominated \citep{pickering97, mcgaugh01}. 
These properties, added to their typically low
metallicities \citep{denaray04, gerritsen99}, lead to the inference that 
even large LSB galaxies are under-evolved
compared to their high surface brightness (HSB) counterparts.  Once their
typically low gas surface densities (M$_{HI} \le 10^{21}\;\rm cm^{-2}$) \citep{pickering97}
and low baryonic-to-dark matter ratios \citep{gurovich04, mcgaugh00}
are taken into account, the question becomes less
why LSB galaxies are under-evolved than how they can form stars at all
\citep[and references therein]{oneil00}.  Yet large LSB galaxies have the same
total luminosity within them as ordinary Hubble 
sequence spirals
\citep{oneil04, impey97, pickering97, sprayberry95}.  
On average then, star formation cannot be too inefficient in these large LSB galaxies
in spite of their unevolved characteristics, 
else their integrated light would be significantly less then in their HSB counterparts.

In an effort to better understand this enigmatic group of galaxies and
their evolutionary status, we recently conducted a 21-cm survey to discover a
larger nearby sample of such objects \citep{oneil06, oneil04}.  We succeeded in
identifying about 25 candidates within the redshift range $0.04 < z <
0.08$, whose combined HI and optical properties suggest them to be large
LSB galaxies.  We obtained B, R, and \Ha\ imaging of 19 of these
galaxies at Lowell Observatory to confirm whether these candidates are
indeed LSB galaxies, and to obtain a dataset of their fundamental parameters.
These observations are presented here, and interestingly, none of the
galaxies ultimately turned out to be LSB galaxies by the strict conventional
definition; we discuss this result below in \S ~\ref{sec:sb}.  However, these galaxies
still represent a sample whose surface brightnesses are
below average, and whose properties are intermediate between those of
the {\it bona fide} massive LSB galaxies, and ordinary HSB galaxies.
In this work, we quantify and parameterize the fundamental properties
of this sample of large, ``lower surface brightness'' galaxies.

\section{Galaxy Sample}
\label{sec:sample}

There are three ways that disk galaxy surface brightness can be measured or
quantified --  using a surface brightness profile and fitting an
exponential disk to derive the central surface brightness; 
measuring an average surface brightness within a given isophotal diameter;
and measuring the surface brightness of the isophote at
the 1/2 light radius point (the effective surface brightness).  
The latter two definitions suffer from the fact that the bulge light
is included in the surface brightness estimates, resulting in their 
prediction of the {\it disk} surface brightness to be less accurate.
As a result, the typical operational definition of an LSB galaxy uses the first definition,
and defines an LSB galaxy as one whose whose observed disk central surface brightness
is $\mu_B(0) \ge$23.0 \mss.  For reference, the Freeman value of 
$\mu_B(0)$=21.65 +/- 0.30 \mss\ defines the distribution of central
surface brightness, in the blue band, for Hubble sequence spirals. 

Regardless of the definition, without pre-existing high quality
optical imaging of galaxies, it is difficult to unambiguously identify a sample of disk
galaxies that will turn out to be LSB.    With only catalog data available, one is driven to
use the average surface brightness and identify potential LSB galaxies as those whose average 
surface brightness is below some threshold level.

All of the galaxies in this sample were identified as LSB by \citet{bothun85}
using the magnitude and diameter values found in the {\it Uppsala General Catalog} \citep{nilson},
and employing the general equation $\langle \mu_B \rangle = m_{pg}+5log(D) + 8.63$.
Here, $m_{pg}$ is the photographic magnitude of the galaxies, D is the diameter in
arcminutes, and the constant, 8.63, is derived from the conversion from arcminutes to arcseconds
(8.89) and the conversion from m$_{pg}$ to m$_B$ (-0.26, as used by Bothun, \etal)
\citet{bothun85} then made a cut-off to the galaxies in
their sample, requiring $\langle \mu_B \rangle >$24.0 mag arcsec$^{-2}$ to look for galaxies
with lower surface brightness disks, with the majority of the galaxies chosen having 
$\langle \mu_B \rangle >$25.0 mag arcsec$^{-2}$ .  (The inclusion of a number of galaxies with 
 $\langle \mu_B \rangle$=24-25 mag arcsec$^{-2}$ was due to the 0.5mag errors given in the
UGC.)

The \citet{bothun85} sample was further pared down by our desire to image large
LSB galaxies.  That is, we wished to avoid the dwarf galaxy category entirely.
To do this, we required the galaxies to have
M$_{HI}>10^{9}$ \Msol, W$_{20} >$ 200 km s$^{-1}$, and/or M$_B < -19$. 
These criteria are sufficiently removed from the dwarf galaxy category to guarantee
no overlap between our sample and that category exists.

\section{Observations \& Data Reduction}
\label{sec:reduce}
Galaxies' integrated broad-band colors represent a convolution of the
mean age of the stellar population, metallicity, and recent star formation rate;
while measurements of \Ha\ luminosity provide a direct measure of the
current star formation rate (SFR).  With these combined observations,
is is possible to parameterize the current SFR relative to the
overall star formation history.  As a result, these observations are
widely used in many surveys that target fundamental galaxy parameters,
for example, SINGG \citep{meurer06} and 11HUGS \citep{kennicutt04},
and others \citep[e.g.][]{gavazzi06, koopman06, helmboldt05}.  

19 galaxies were observed on 7-10 June, 2002 and 5-8 October, 2003 using the Lowell 
1.8m Perkins telescope.  The filter set used included Johnson B and R as well as 
three \Ha\ filters from a private set (R. Walterbos) with center frequency/bandwidths
of 6650/75, 6720/35, 6760/75 \AA.  A 1065x1024 pixel Loral SN1259 
CCD camera was used, giving a  3.3$^\prime$ field of view and resolution of 0.196\acs/pixel.  
Seeing in June, 2002, ranged from 1.8\acs\ - 2.4\acs\ and from 1.4\acs\ - 2.2\acs\ for 
the October, 2003, observations.  At least 3 frames, each shifted slightly in position, 
were obtained for each object through each filter and were median filtered to reduce 
the effect from cosmic rays, bad pixels, etc.  All initial data reduction (bias 
and flat field removal, image alignment, etc) was done within IRAF.  The R band 
images were scaled and used as the continuum images for data reduction purposes.

Corrections to the measured fluxes were made in the following way.
Atmospheric extinction was obtained using the observational airmass and
the atmospheric extinction coefficients for  Kitt Peak which are distributed with IRAF.  
Galactic extinction was corrected using the values for E(B$-$V) obtained
from NED, the reddening law of \citet{seaton79} as parameterized by \citet{howarth83}
($A(\lambda)=X(\lambda) E(B-V)$) and assuming the case B recombination of \citet{osterbock89} with
R$_V$=3.1 \citep{odonnell94} (X(6563\AA)=2.468).
Contamination from $\left[{NII}\right]$ emission in the \Ha\ images was corrected using the relationship derived by
\citet{jansen00} and re-confirmed by \citet{helmboldt04}: 
\[ log{{\left[NII\right]}\over{H\alpha}}=\left[-0.13\pm 0.035\right]M_R + \left[-3.2\pm 0.90\right], \]
where $M_R$ is the absolute magnitude in the R band.  \Ha\ extinction was determined using the 
equation found in \citet{helmboldt04}:
\[log\left( H\alpha \right)_{int} = \left[ -0.12 \pm 0.048\right] M_R + \left[ -2.5 \pm 0.96\right] \]
which was found through a linear least squares fitting to the $A(H\alpha)_{int}$ determined
using all galaxies in his sample with a measured H$\beta$ flux.  For this calculation, \citet{helmboldt04}
used the \Ha\ to H$\beta$ ratio measured by \citet{jansen00}, an assumed intrinsic ratio of
${{H\alpha}\over{H\beta}}$=2.85 (Case B recombination and T=10$^4$ K \citep{osterbock89}), the extinction curve
of \citet{odonnell94}, and R$_V$=3.1
No correction for internal extinction due to inclination was made for the B and R bands.
It should be noted, though, that in a number of plots
inclination corrections were made to the B and R colors and central surface brightnesses, as noted in the 
Figure captions.  The corrections used in
these cases are:
\begin{equation} \mu(0)^\lambda _{corr} = \mu(0)^\lambda - 2.5C^\lambda log(b/a) \label{eqn:mu_inc}\end{equation}
and
\begin{equation} m^\lambda _{corr} = m^\lambda - A^\lambda \label{eqn:mag_inc_1}\end{equation}
\begin{equation} A^\lambda = -2.5log\left[f\left(1 + e^{-\tau^\lambda sec(i)}\right)
 \:+\;\left(1 - 2f\right) \left({{1 - e^{-\tau^\lambda sec(i)}}\over{\tau^\lambda sec(i)}}\right) \right]
\label{eqn:mag_inc_2}\end{equation}
Here, C$^{R,B}$=1 \citep{verheijen97}; $\left(b/a\right)$ is the ratio of the minor to major axis;
f = 0.1 and $\tau_{R,B}$=0.40, 0.81 \citep{tully98,verheijen97}.   Finally, a correction was applied
to account for the effect of stellar absorption in the Balmer line of
\begin{equation} F_{cor} = F_{obs}\left({1 + {{W_a}\over{W_e}}}\right),\end{equation}
where $F_{cor}$ is the corrected and $F_{obs}$ is the observed \Ha\ flux, W$_e$ is the measured 
equivalent width and W$_a$ is the equivalent width of the Balmer absorption lines.  As we do not 
have measurements for W$_a$, we estimated W$_a$ to be 3$\pm$1 \AA, based off the values found in 
\citet{oey93, ronnback95, mccall85}.  Note that this effect is potentially stronger in the diffuse
gas than in the \ion{H}{2} regions due to the older stellar population likely lying in the diffuse gas.
As a result we may still be underestimating the total \Ha\ flux in the diffuse gas within the galaxies.
However, as the diffuse gas fractions for these galaxies are extremely high 
(see Section~\ref{sec:StarFormation}, below), it is unlikely that this effect is high.

\begin{figure}
\centerline{
\includegraphics[width=2.5in]
{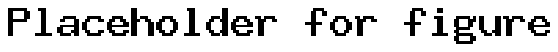}}
\caption{Grey scale images of the observed galaxies. Figure available through the published AJ paper or online at http://www.gb.nrao.edu/$\sim$koneil. \label{fig:morph}}
\end{figure}

\begin{figure}
\centerline{
\includegraphics[width=2.5in]
{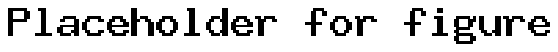}}
\caption{Surface brightness profiles for all galaxies observed. The dash-dotted lines show the inner fit, the dashed lines show
the outer fit, and the solid lines show the combined fits.  Both the B (blue - bottom) and R (red - top) profiles are shown.  Figure available through the published AJ paper or online at http://www.gb.nrao.edu/$\sim$koneil.\label{fig:fits}}
\end{figure}

\begin{figure}
\centerline{
\includegraphics[width=2.5in]
{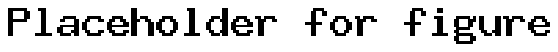}}
\caption{Color profiles for all the galaxies observed. Here the inner fits (when made) are shown by a dashed line and the outer fits are 
shown by a solid line.  Figure available through the published AJ paper or online at http://www.gb.nrao.edu/$\sim$koneil.\label{fig:colors}}
\end{figure}

Global parameters and radial profiles for the galaxies were determined primarily using the routines 
available in IRAF (notably {\it ellipse}) and the results are given in Tables~\ref{tab:global} and \ref{tab:global2}.  
Galaxy images, surface brightness profiles, and color profiles are given in 
Figures~\ref{fig:morph} -- \ref{fig:colors}.  In all cases the inclination and
position angle for the galaxies were determined from the best fit values from the 
B \& R frames.  These best fit values were then used for the ellipse fitting in 
all four images (B, R, \Ha\, and continuum with \Ha\ subtracted), a practice which insures the 
color profiles are obtained accurately and are not affected
by, e.g. misaligned  ellipses.  The same apertures were also used for all four images,
with the apertures found through allowing {\it ellipse} to range from 1 pixel (0.196$^{\prime\prime}$)
until the mean value in the ellipse reaches the sky value, increasing geometrically by a factor
of 1.2.  Sky values were found through determining the mean value in more than 100 5$\times$5 sq. pixel boxes
in each frame.
The error found for the sky was incorporated into all magnitude and surface brightness errors, 
which also include errors from the determination of the zeropoint and the errors from the 
\ion{N}{2} contribution to the \Ha\ (in the case of the \Ha\ magnitudes).

The B and R surface brightness profiles of all galaxies were fit using two methods.  
First, the inner regions of the galaxies' surface brightness profiles was fit
using the de Vaucouleurs r$^{1/4}$ profile
\begin{equation} \Sigma(r) = \Sigma_{eff} exp^{-7.669\left[{(r/r_{eff})^{1/4} - 1}\right]} \rightarrow 
\mu(r) = \mu_{eff} + 8.327\left[{\left({{r}\over{r_{eff}}}\right)}^{1/4} - 1\right], \label{eqn:bulge}\end{equation}
and the outer regions were fit by the exponential disk profiles
\begin{equation} \Sigma(r) = \Sigma_0 exp^{\left({{r}\over{\alpha}}\right)} \rightarrow
\mu(r) = \mu_0 + 1.086 \left({{r}\over{\alpha}}\right).\label{eqn:disk} \end{equation}
Additionally, we attempted to fit a disk profile (\ref{eqn:disk}) to both the inner and outer regions
of the galaxies', to determine if a two-disk fit would better match the data 
\citep{broeils97,dejong96}.
Roughly one-fourth of the galaxies (5/19) were best fit (in the $\chi^2$-sense) by the 
standard bulge+disk model.  Another 47\% of the galaxies were best fit by the two-disk
model.  Of the remaining galaxies, 21\% (4 galaxies) were best fit by a single disk, and
one galaxy (UGC~11840) could not be fit by any profile.  The results from the fits are
shown in Table~\ref{tab:fitted} and Figure~\ref{fig:fits}, and an asterisk (*) is placed
next to the best fit model.
Note that in a few cases (e.g. UGC~00189) only one model is listed in the
Table.  This is due to the fact that in these cases the fitting using the other model proved to be 
completely unrealistic.
Finally, it should be noted that in all cases the same best-fit model was used for both 
the B and R data.

The color profiles were similarly fit (using an an inverse error weighting)
with a line to both the inner and outer galaxy regions 
(Figure~\ref{fig:colors}).  Here, though, the ``boundary radius" was simply taken from the
surface brightness profile fits, with the ``boundary radius" being defined as the radius where the inner
and outer surface brightness fits crossed.  If only one (or no) fit was made to the surface 
brightness profile, then only one color profile was fit.  In a number of cases the difference
in slope between the inner and outer galaxy regions was less than the least-squares error for the fit. 
In these cases again only one line was fit for the color profiles.

\begin{figure}
\centerline{
\includegraphics[width=2.0in]
{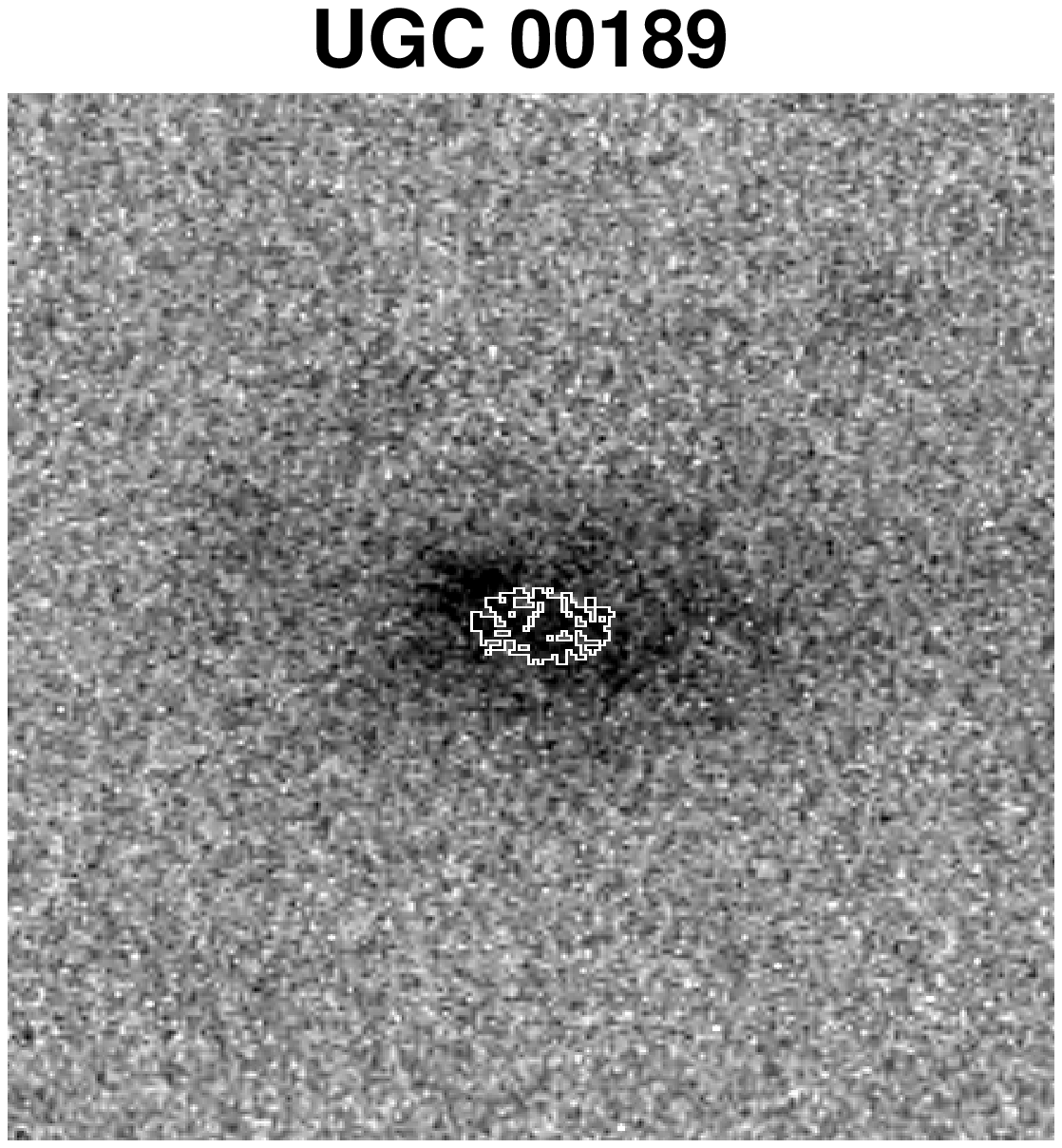}
\includegraphics[width=2.0in]
{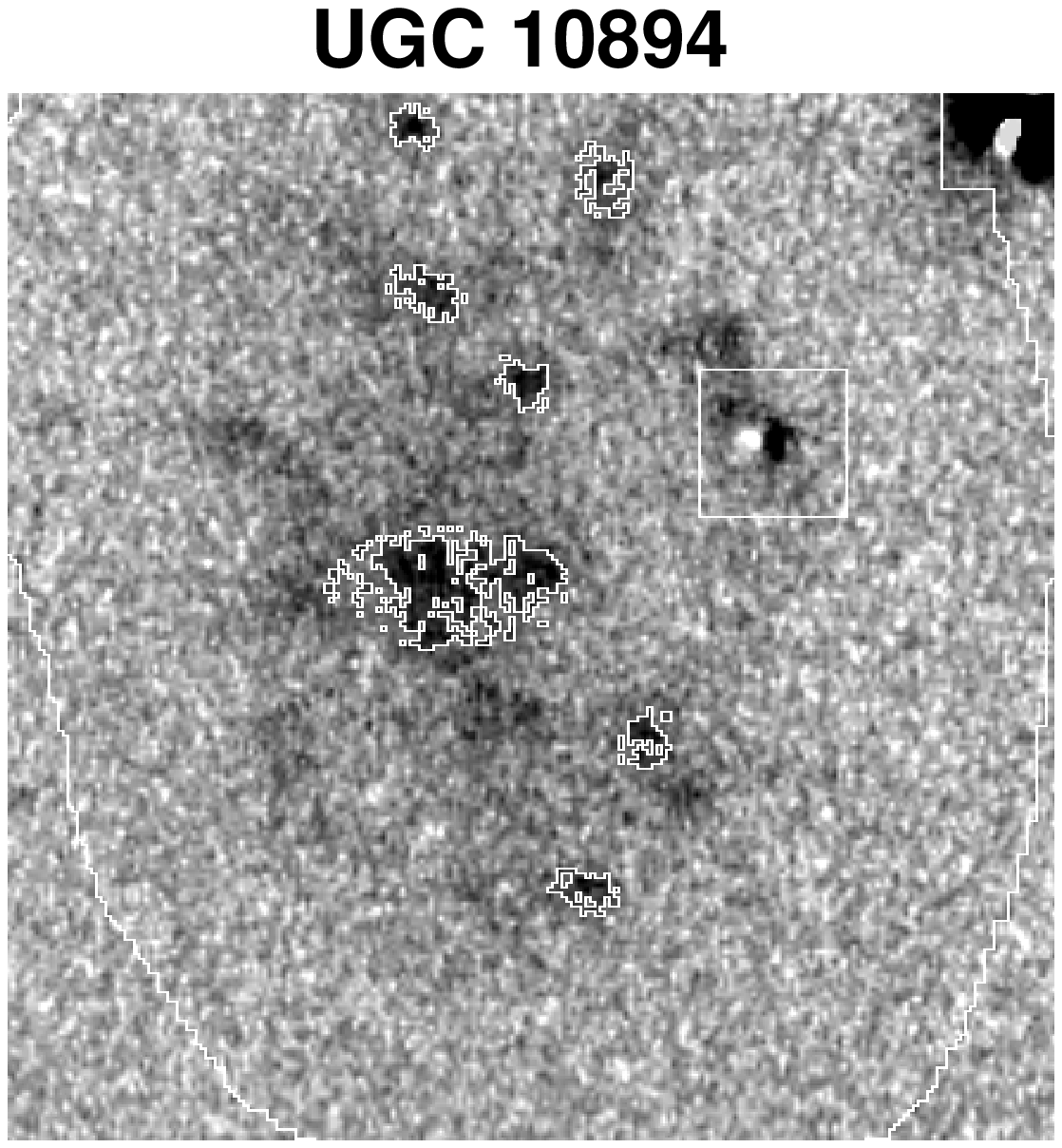}
}\centerline{
\includegraphics[width=2.0in]
{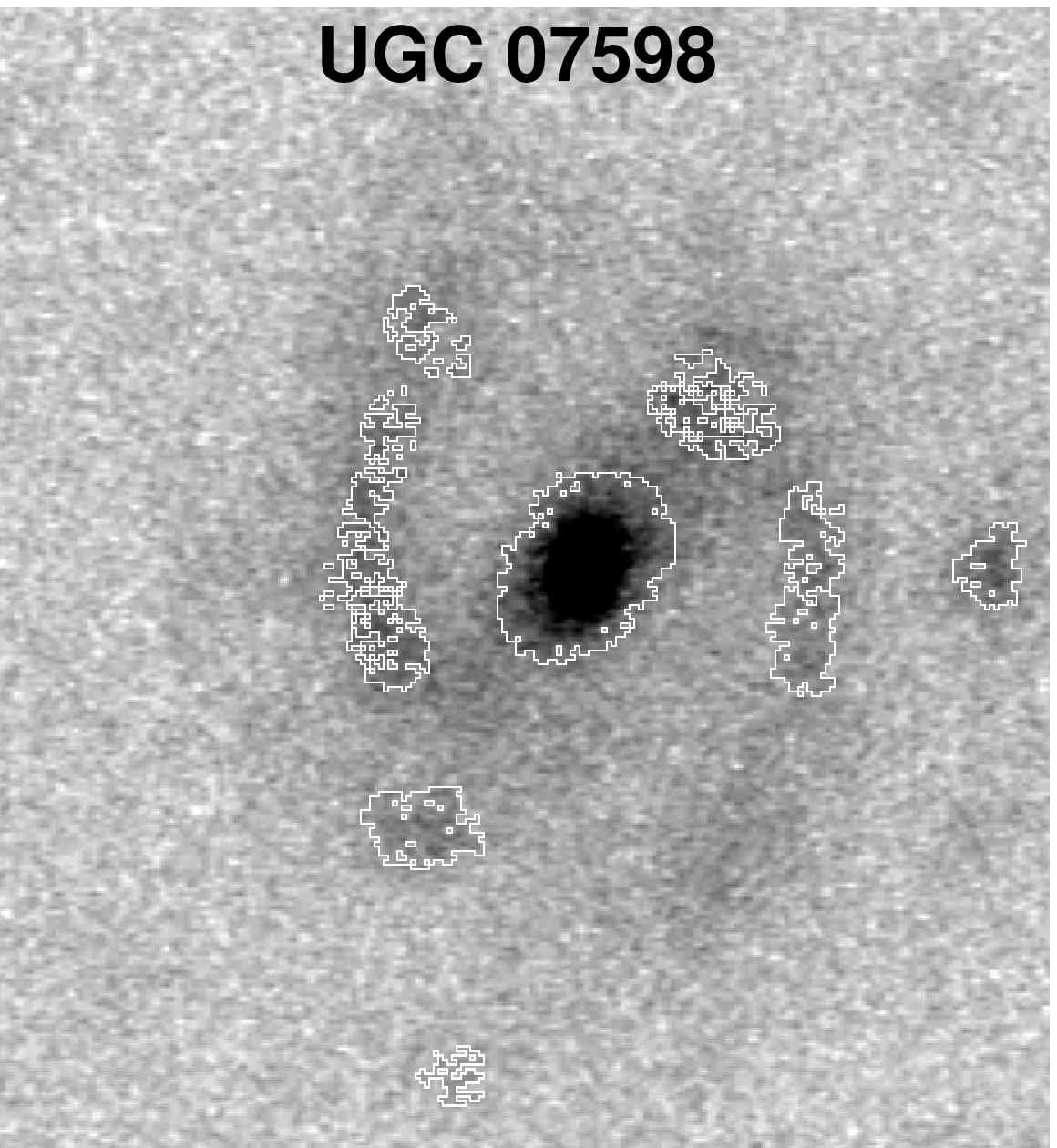}
\includegraphics[width=2.0in]
{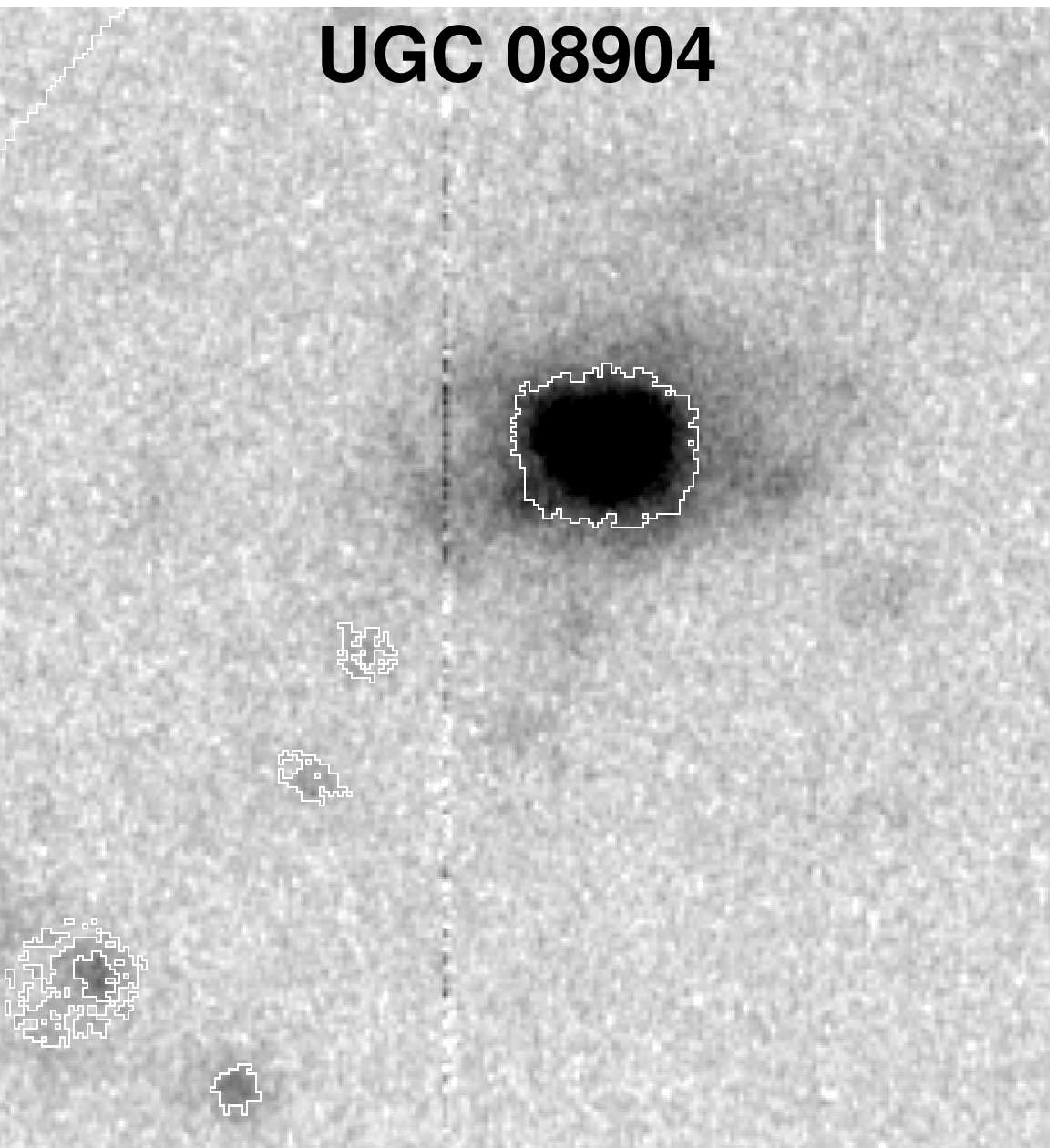}
}
\caption{Example images showing the \ion{H}{2} regions found by HIIphot
for the galaxies UGC~00189, UGC~10894, UGC~07598, and UGC~08904.  The \ion{H}{2}
regions are outlines in white.  In the case of UGC~10894 two regions which were masked due to the presence of stars can also be seen, outlined by the square white boxes. \label{fig:regions}}
\end{figure}

The HIIphot program \citep{thilker00} was used both to determine the shape and 
number of \ion{H}{2} regions for each galaxy and also to determine the \Ha\ flux 
for each of these regions.  
The fluxes from the \Ha, \Ha-subtracted continuum, B, and R images were measured in identical corresponding apertures,
which are
the \ion{H}{2} region boundaries defined by  HIIphot.  While HIIphot applies an interpolation
algorithm across these apertures to estimate the diffuse background in the H$\alpha$
frames, we determined the background in other bands from the median flux in
an annulus around each \ion{H}{2} region aperture.
Results from the analysis of 
the \ion{H}{2} regions are given in Table~\ref{tab:regions}, and sample \ion{H}{2} regions 
are shown in Figure~\ref{fig:regions}.  Errors for the \Ha, SFR, and EW measurements are 
derived from the error values reported with HIIphot.  Errors for the B and R magnitudes, and colors,
are derived from the total sky and zeropoint errors, as well as the error in positioning of the 
HII regions.  The diffuse fraction errors are derived both from the total \Ha\ flux errors and
also include errors in determining the total flux within the HII regions and for the entire galaxy.
Finally, it should be mentioned that the equivalent width (EW) was calculated simply as the 
ratio of the \Ha\ flux to \Ha-subtracted continuum flux for 
a given region (or the whole galaxy). 

The large distances to the observed galaxies (40 - 100 Mpc) results in many of the
\ion{H}{2} region being blended together.  As a result, any luminosity function 
derived for these objects would be necessarily skewed towards larger HII regions \citep[see][]{oey06}.
This can be seen in the analysis done by \citet{thilker00} wherein the dependence of the
luminosity function found for M51 was examined.  There one can clearly see the increase
in the number of high luminosity regions and subsequent reduction in the number of low luminosity
regions as the galaxy is 'moved' to increasing distances.  Examining their results also
shows that while the {\it distribution} of \ion{H}{2} region luminosities changes with distance,
the {\it total} luminosity of the \ion{H}{2} regions, as found by HIIphot, does not change
significantly as the galaxy moves from 10 Mpc to 45 Mpc.  As a result, while determining 
luminosity functions for the galaxies in this paper is not feasible due to the distances involved,
derivations such as the diffuse fraction are unaffected by distance.  This fact is also
supported by the SINGG survey results \citep{oey06}.

\section{Surface Brightness}
\label{sec:sb}

\begin{figure}
\centerline{
\plottwo{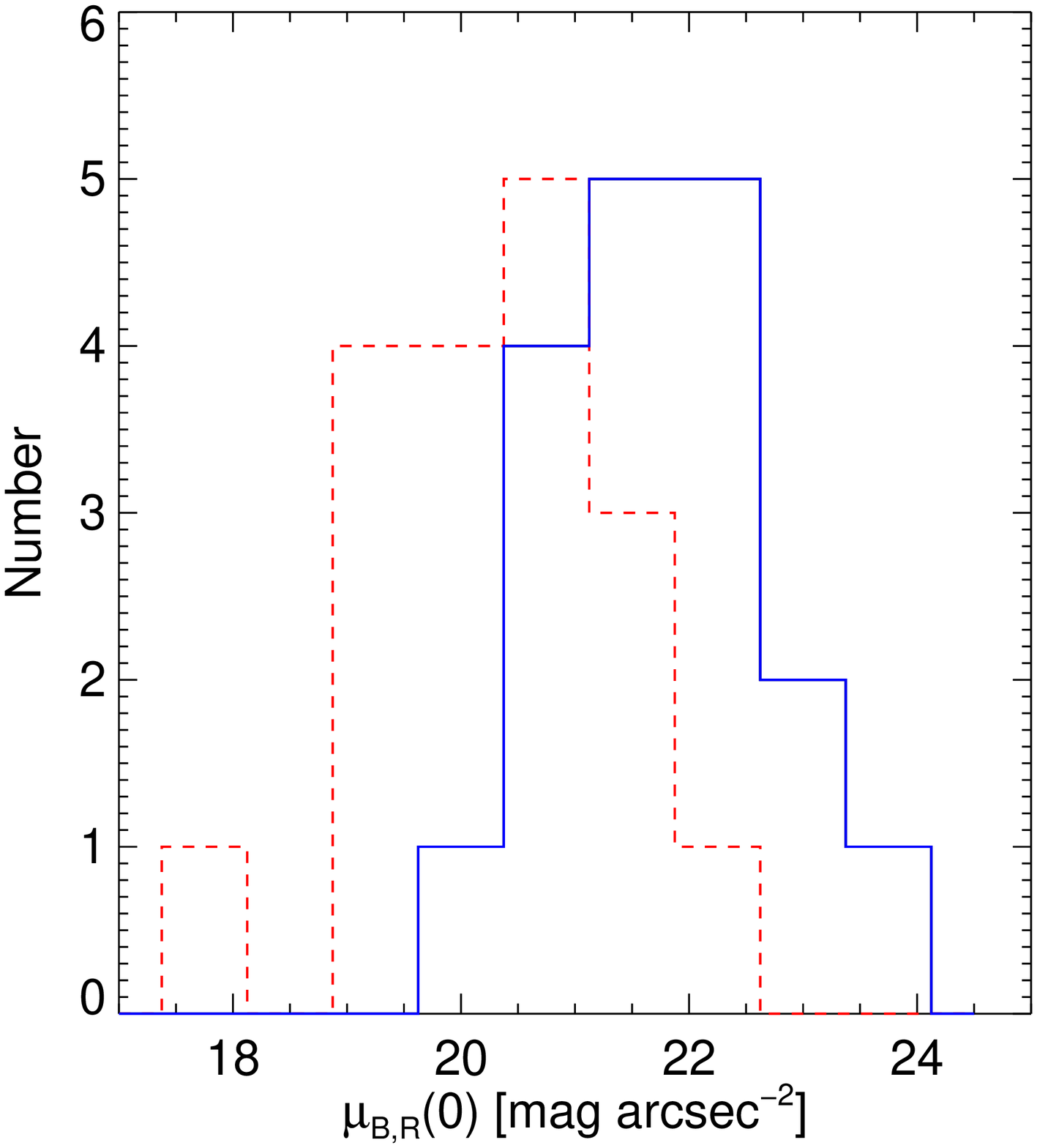}{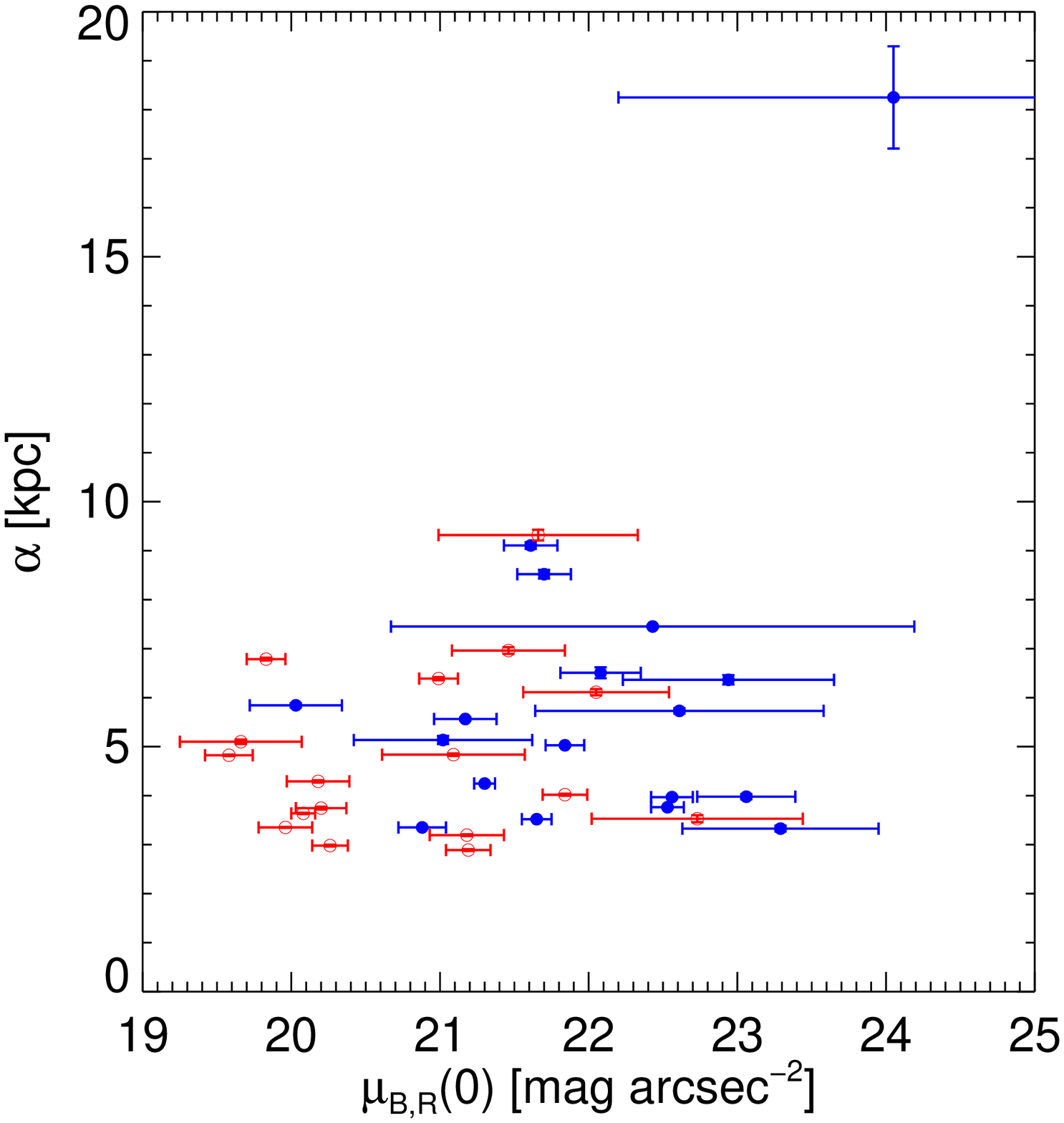}
}
\caption{(Left) Histogram showing the distribution of central surface brightnesses for the observed galaxies.
The (red) dashed line shows the R-band data and the (blue) solid line shows the B-band data. 
(Right) Plot of the observed central surface brightness against scale length of the outer disk.
Here, the R-band data is demarcated by (red) open circles while the B-band data uses (blue) filled circles.
\label{fig:mu_hist}}
\end{figure}

The distribution of central surface brightnesses found for the galaxies observed is shown in Figure~\ref{fig:mu_hist}.
As is plain from that Figure, the mean measured central surface brightness for this sample, falls 
short of the definitions discussed in Section~\ref{sec:sample}.  Indeed only 4 galaxies in our
sample meet the operational definition of LSB galaxies as having $\mu_B(0)\ge$ 23 \mss.
If we return to the Freeman value, however, we see that the operational definition of LSB galaxies
is 4.5$\sigma$ from the value for Hubble sequence spirals,
making it statistically extreme.  For the sample defined here, half have central surface brightnesses
at least two sigma above the Freeman value, a definition only 2.5\% of the Freeman sample meets.
As a result, while the sample does not meet the operation criteria for LSB galaxies, we clearly do have a sample
with lower central surface brightnesses that would be found in the average Hubble sequence galaxies.

It should be pointed out here that the main scientific focus of \citet{bothun85} was not oriented toward producing
a representative sample of LSB galaxies as detected on photographic surveys 
\citep[that focus did not occur until ][]{schombert88}, but rather toward identifying cataloged galaxies for 21-cm based
redshift determinations.  The galaxies were chosen to have surface brightnesses that were too low
for reliable optical spectroscopy (assuming emission lines were not present).  This was done as a test of
the potentially large problem of bias in on going optical redshift
surveys in the time \citep[see][]{bothun86}.  In fact, the operational criteria for
selecting the galaxies that were observed at Arecibo 20 years ago, lay in the knowledge that these
cataloged galaxies were never going to be even attempted in the
optical redshift surveys of the time and this raised the very real possibility of biased redshift
distributions and an erroneous mapping of large scale structure.
 
In the original redshift measurements of \citet{bothun85} a significant number of candidate LSB
galaxies were not detected at 21-cm within the observational redshift window (approximately 0-12,000 km/s).
Many of those non-detections would later turn out to be intrinsically large galaxies located at redshifts
beyond 12,000 km/s \citep[see][]{oneil04}.   As we are interested here in the \Ha\ properties of galaxies
with large, relatively LSB disks, these initial non-detections comprise the bulk of our sample. 

Surface photometry of this sample not only provides detailed information regarding
the galaxies' surface brightness and color distributions, but it also probes the 
efficacy of the \citet{bothun85} average surface brightness criteria for selecting LSB disks.
Here, we used the magnitudes and diameters obtained in this study (Table~\ref{tab:global})
with two different equations for 
determining a galaxy's average surface brightness within the D$_{25}$ radius. 
The first equation used is that of \citet{bothun85}
\begin{equation} \langle \mu_{25} \rangle=m_{25}+5log(D_{25})+3.63 \label{eqn:bothun}\end{equation}
and the second is a modified version of the above equation from \citet{bottinelli95} 
which takes the galaxies' inclination into account:
\begin{equation}\langle \mu_{25} \rangle
=m_{25}+5log(D_{25})+3.63-2.5log\left[kR^{-2C}+\left(1-k\right)R^{(0.4C/K)-1}\right].\label{eqn:bottinelli}\end{equation}
In both equations, m$_{25}$ and D$_{25}$ are the magnitude and diameter (in units of 0.1$^\prime$) at
the $\mu$=25.0 \mss\ isophote, R is the axis ratio (a/b), and C is defined as (logD/logR) and is fixed at 0.04 
\citep{bottinelli95}.  Finally, k (the ratio of the bulge-to-disk luminosity) and 
K (a measure of how the apparent diameter changes with surface brightness at a given axis ratio)
are dependent on the revised de Vaucouleurs morphological type (T) as follows \citep{simien86,fouque85}:\\
T=1 $\rightarrow$ k=0.41;
T=2 $\rightarrow$ k=0.32;
T=3 $\rightarrow$ k=0.24;
T=4 $\rightarrow$ k=0.16;
T=5 $\rightarrow$ k=0.09;
T=6 $\rightarrow$ k=0.05;
T=7 $\rightarrow$ k=0.02;
T$\ge$8 $\rightarrow$ k=0.0; \\
\[K=0.12-0.007T\:\; if\:\; T<0;\: K=0.094\:\; if\:\; T \ge 0. \]
The values for k at T$\ge$8 are extrapolated from fitting the \citet{simien86} values.

\begin{figure}
\centerline{
\plottwo{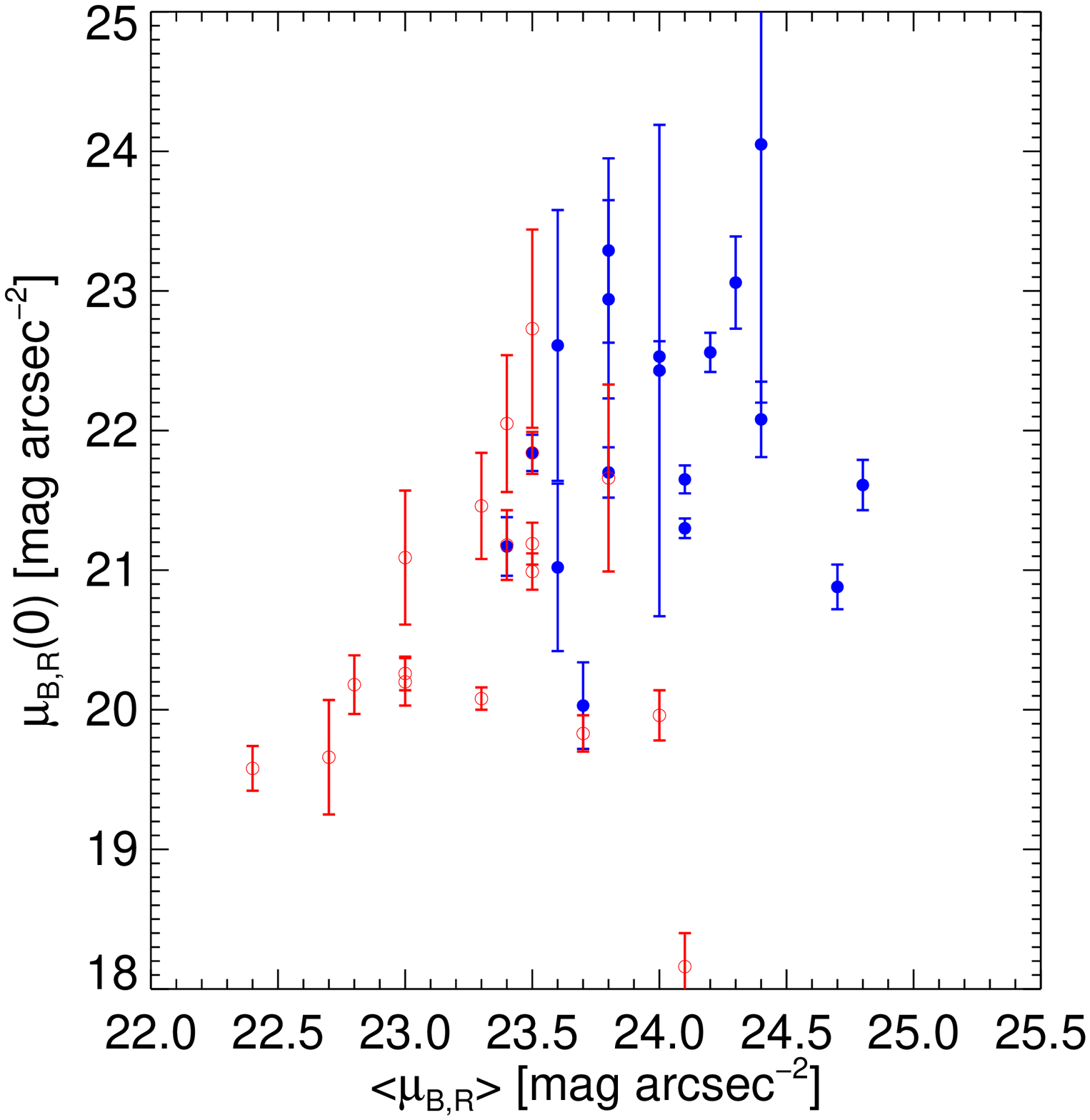}{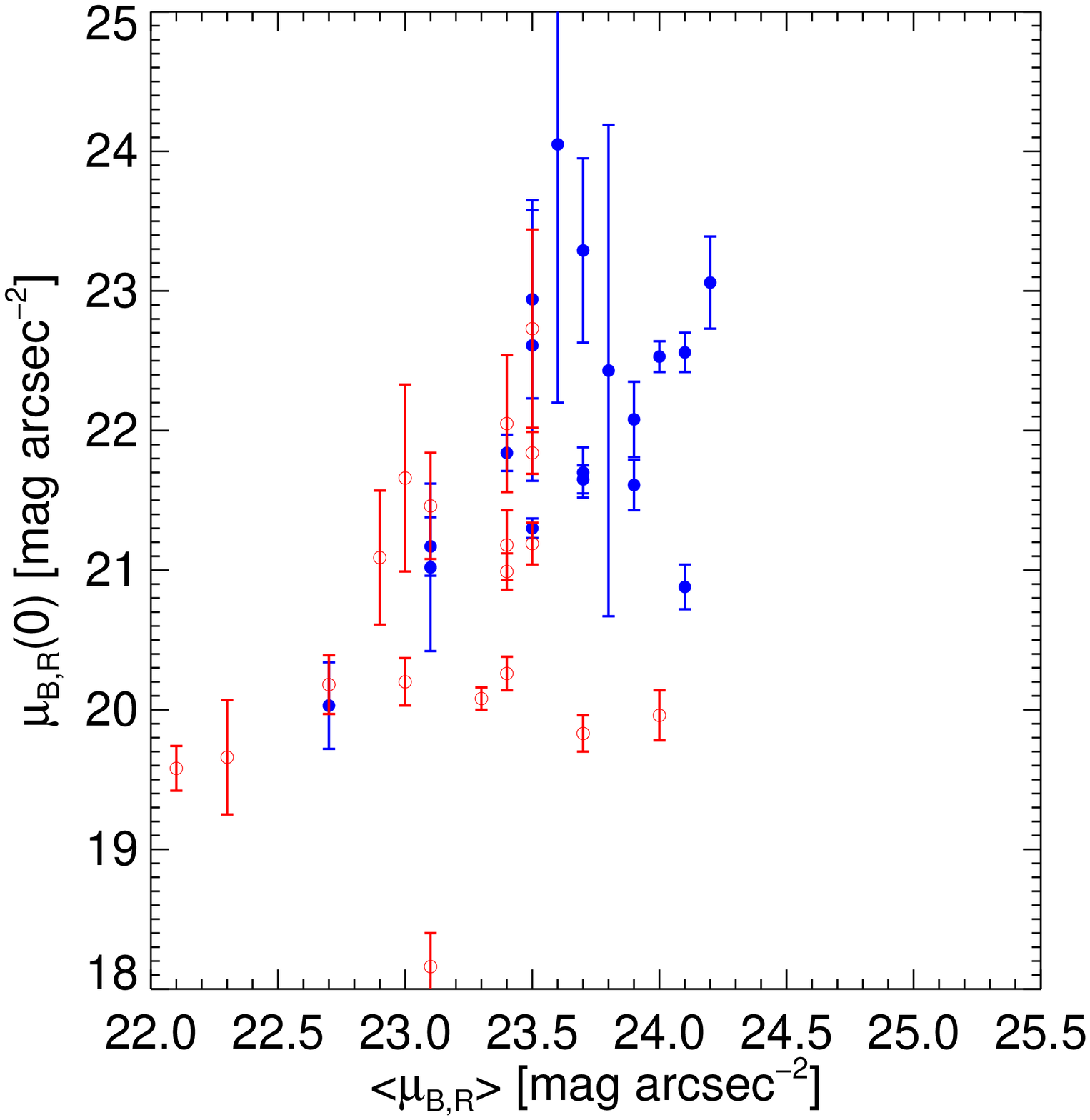}
}
\caption{Plots comparing the measured central surface brightnesses with the average surface brightness for the galaxies, as 
defined by Equation~\ref{eqn:bothun} (left) and Equation~\ref{eqn:bottinelli} (right) 
and using the magnitude and diameter values 
found herein.  The R-band data is demarcated by (red) open circles while the B-band data uses (blue) filled circles. 
\label{fig:mu_mu}}
\end{figure}

\begin{figure}
\centerline{
\plottwo{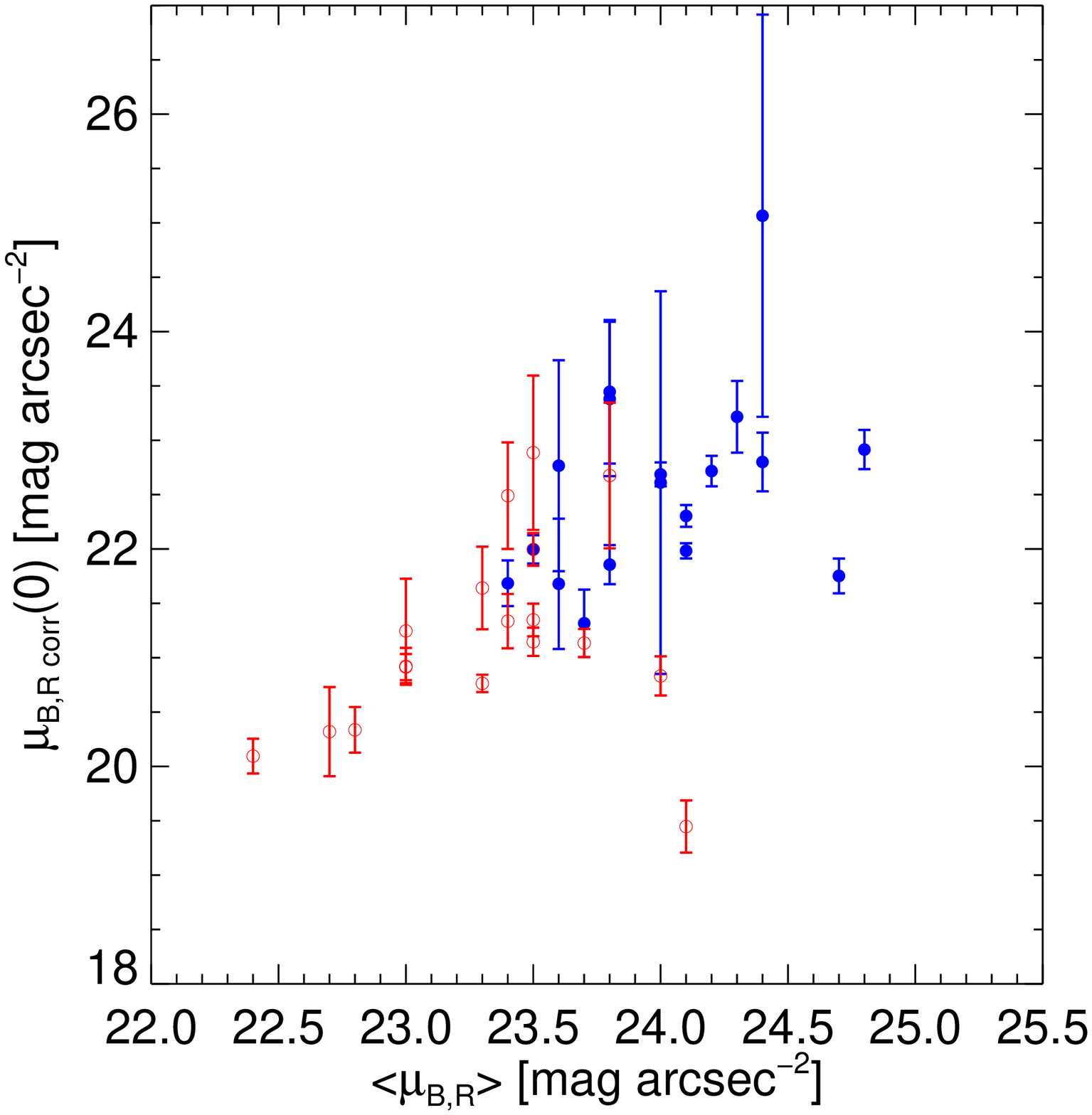}{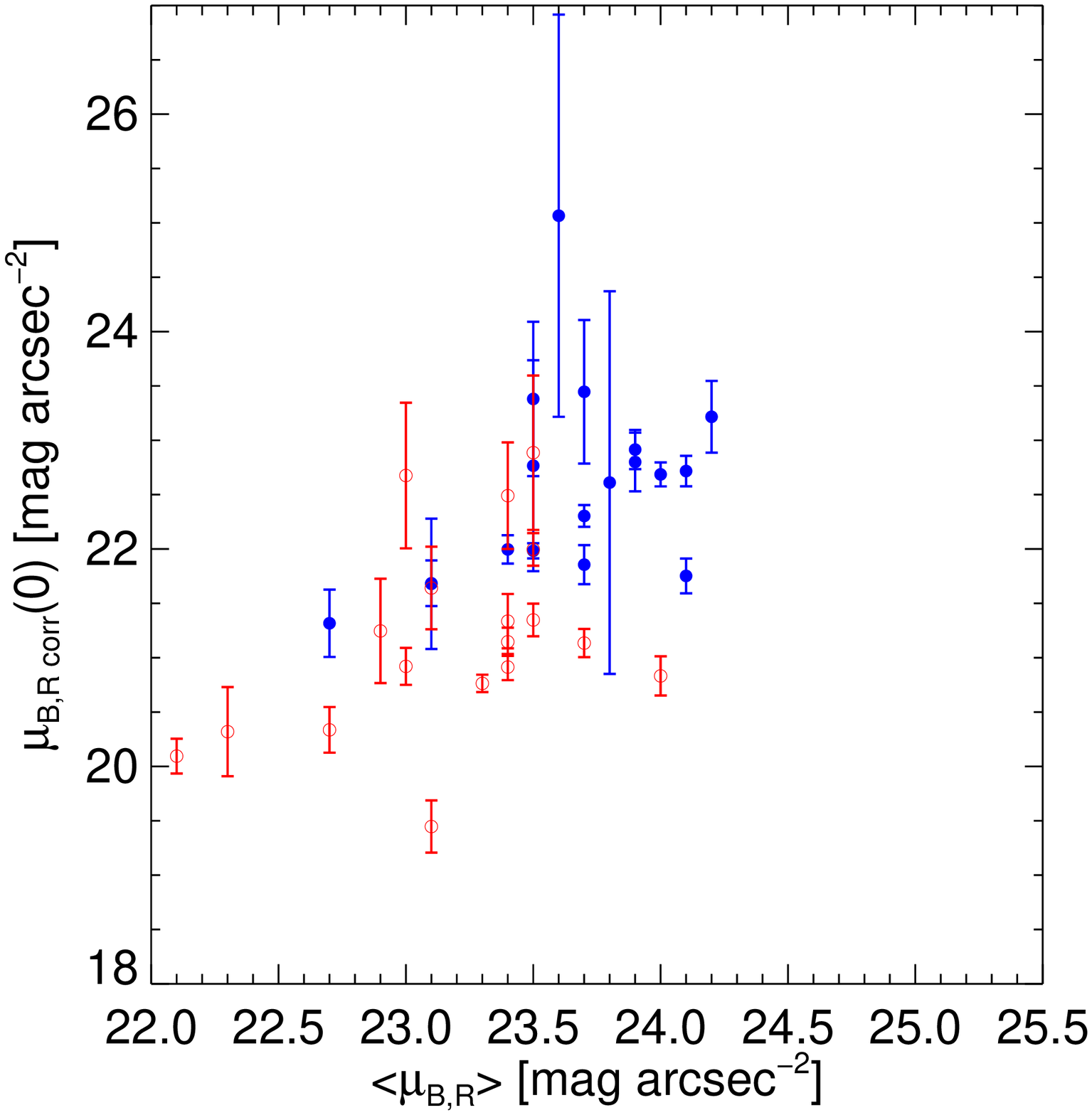}
}
\caption{Plots comparing the measured central surface brightnesses, corrected for inclination,
with the average surface brightness for the galaxies, as
defined by Equation~\ref{eqn:bothun} (left) and Equation~\ref{eqn:bottinelli} (right)
and using the magnitude and diameter values
found herein.  The R-band data is demarcated by (red) open circles while the B-band data uses (blue) filled circles.
\label{fig:mu_mu_inc}}
\end{figure}

The results of equations \ref{eqn:bothun} and \ref{eqn:bottinelli}, plotted against the galaxies' central
surface brightness both uncorrected and corrected for inclination, are shown in
Figures~\ref{fig:mu_mu} and \ref{fig:mu_mu_inc}, respectively.  The difference between the two plots
is small, with neither equation doing an excellent job in predicting when a disk's central surface brightness
will be low.  The two equations (\citet{bothun85} and \citet{bottinelli95}) have roughly the same fit (in the 
$\chi^2$ sense), which at first appears surprising. 
It is likely that uncertainties in the inclination measurements
and morphological classification of the galaxies have increased the scatter in the \citet{bottinelli95}
equation, increasing the scatter in an otherwise more accurate equation.  
As a result, while the  \citet{bottinelli95} may indeed be the most accurate,  
the simpler equation is equally as good to use in most circumstances as it involves fewer assumptions.

The second fact that is readily apparent in looking at Figures~\ref{fig:mu_mu} and \ref{fig:mu_mu_inc}
is that with the new measurements of magnitude and diameter, {\it none} of the galaxies in our sample
meet the criterion laid out by \citet{bothun85} for an LSB galaxy.  That is that none of the galaxies
in this sample have $\langle \mu_{25} \rangle >$25 \mss.  As \citet{bothun85} listed all of these objects
as having $\langle \mu_{25} \rangle\;>$25 \mss\ using the magnitudes and diameters provided by the original
UGC measurements, this shows that the UGC measurements indeed predicted fainter magnitudes/larger values for
D$_{25}$ than is found with more sophisticated measurement techniques.  Additionally, it is good to note
that the trends shown in Figures~\ref{fig:mu_mu} and \ref{fig:mu_mu_inc} indicate that any galaxy which
met the $\langle \mu_{25} \rangle >$25 \mss\ criteria would be highly likely to also have $\mu(0) >$23 \mss.

In these days of digital sky surveys it is difficult to appreciate the
immense undertaking that defines the UGC catalog.  Anyone who has
looked at the Nilson selected galaxies on the Palomar Observatory Sky Survey (POSS) plates
with a magnifying eyepiece really has to marvel that Nilson's eye saw objects at least one
arcminute in diameter.  It is thus not surprising that, at the ragged
end of that catalog, many of the listed UGC diameters are systematically high.
\citet{cornell87} made a detailed diameter comparison between diameters as obtained
from high quality CCD surface photometry and the estimates made by \citet{nilson}.
They compared the diameter at the 25.0 mag arcsec$^{-2}$ isophote in CCD B images 
to the tabulated diameter in the UGC.  The study, based on approximately 250 galaxies,  identified two
sources of systematic error (neither of which are surprising).  First,  galaxies with reported diameters 
less than 2$^\prime$ typically had D$_{25,B}$ as measured by the CCD images that were 15-25\% smaller.
Second, \citet{cornell87} found a systematic bias as a function of surface brightness in the sense that
lower surface brightness galaxies had a higher number of overestimated diameters in the UGC
than higher surface brightness galaxies. 
It should also be noted that the majority of the galaxies in this study lie at low Galactic latitude.  This seems to be a perverse
consequence that there is a large collection of galaxies between 7,000 -- 10,000 \kms\ (where the diameter criterion
in the UGC yields a relatively large physical size) located at relatively low galactic latitude.
Nominal corrections
for galactic extinction made by \citet{bothun85} turned out to underestimate the extinction as shown
by later published extinction maps.   In some cases, the differences were as large as one magnitude.
The combination of these facts with the very uncertain 
magnitudes of many of these galaxies \citep[see][]{bothun90}, it is not surprising that the measured
average surface brightness could easily be 1-1.5 magnitudes higher than the average surface brightness
that has been estimated from the UGC catalog parameters (roughly 40\% of this comes from systematic
magnitude errors and 60\% from the diameter errors discovered by \citet{cornell87}).

\section{Morphology \& Color}

All of the galaxies observed have large sizes ($3\alpha_B$ = 10 -- 54 kpc),
bright central bulges, and well defined spiral structure 
(Figure~\ref{fig:morph}).  In most cases the galaxies can be described as late-type systems
(Sbc and later).  There are, though, a number of exceptions to this rule.
Three of the galaxies, UGC~00023, UGC~07598, and UGC~11355 (Sb, Sc, and Sb galaxies, respectively) have clear nuclear bars.
UGC~08311, classified as an Sbc galaxy, is clearly in the late stages of merging with another system.  In this
case the LSB classification of the galaxy is likely bogus, as the apparently LSB disk is
likely just the remnant the merging process and will disappear as the galaxy compacts after
the merging process.  UGC~8904 is given a morphological type of S? with both NED and HYPERLEDA,
yet the faint spiral arms surrounding it indicate its should be properly classified as an Sbc system.   
UGC~12021 is, like UGC~00023, listed as an Sb galaxy.  Finally, UGC~11068 has a faint nuclear ring
which is most readily visible in the B image.

The differences between the galaxies becomes more apparent when the \Ha\ images are examined.
\citet{hodge83} classify the radial distribution of \ion{H}{2} regions in spiral galaxies into three 
broad categories -- galaxies with \ion{H}{2} region surface densities which 
decrease with increasing radius, galaxies with oscillating \ion{H}{2} region surface densities, 
and galaxies with ring-like \ion{H}{2} density distributions.  To these categories 
we would add a fourth, to include those galaxies with no detectable \ion{H}{2} regions.

The first category of \citet{hodge83} is also the most common, as it includes all galaxies with
generally decreasing radial densities of \Ha.  In the \citet{hodge83} sample this category is dominated 
by Sc -- Sm galaxies but contains all Hubble types.  In our sample, this category includes both 
galaxies with and without significant \Ha\ emission in the spiral arm regions.  This group includes 
UGC~00023, UGC~00189, UGC~02588, UGC~02796, UGC~03119, UGC~03308, UGC~07598, 
and UGC~12021.  Interestingly, of the galaxies listed above, 4/8 are Sb/Sbc galaxies and 3/8 are Sc-Sm
galaxies.  (The last galaxy, UGC~02588, is an irregular galaxy.)

The second category of \citet{hodge83}, galaxies with oscillating densities, is dominated in their 
sample of Sb galaxies. Only a few of the galaxies in this sample fall into this category, 80\% of
which are also Sb/Sc galaxies.  These are 
UGC~02299, UGC~08311, UGC~08904, UGC~11355, and UGC~11396.  These galaxies all have a concentration 
of star formation seen in the nuclear regions and then clumps of star formation spread through the spiral 
arms, typically accompanied by diffuse \Ha\ also spread throughout the arms.  

The third category of \citet{hodge83} is dominated by early-type galaxies, of which we have none in our sample.
Nonetheless we have three galaxies which fall into this category -- UGC~08644, UGC~10894, and UGC~11617.
All three have \ion{H}{2} regions spread throughout their
disks, with no central concentration near the galaxies' nuclei.  In fact, the three brightest star forming
regions within UGC~08644 all lie with the spiral arms, and are visible in all three filters.  In contrast,
both UGC~11617 and UGC~10894 have no bright \ion{H}{2} regions, but instead have a large number of diffuse
\ion{H}{2} regions, with the brightest (as listed in Table~\ref{tab:global2}) receiving that designation simply due to its size.

The fourth category of galaxies contains UGC~01362, UGC~11068, and UGC~11840, none of which have detectable \Ha.
In the case of UGC~01362 and UGC~11840 this is not too surprising as the galaxies are dominated by a bright nucleus,
and their surrounding spiral arms are extremely faint in both R and B.  As a result, 
any \Ha\ which may exist in the galaxies' disks is too diffuse to be detected.
UGC~11068, though, has both a well defined nucleus and a clear spiral structure extending out to a radius of
$\sim$13 kpc (3$\alpha$).  Yet no \Ha\ can be detected in this galaxy.  This may mean that 
UGC~11068 is in a transition state for its star formation, with no ongoing star formation yet with enough
recent activity that the spiral arms remain well defined.  

\begin{figure}
\plottwo{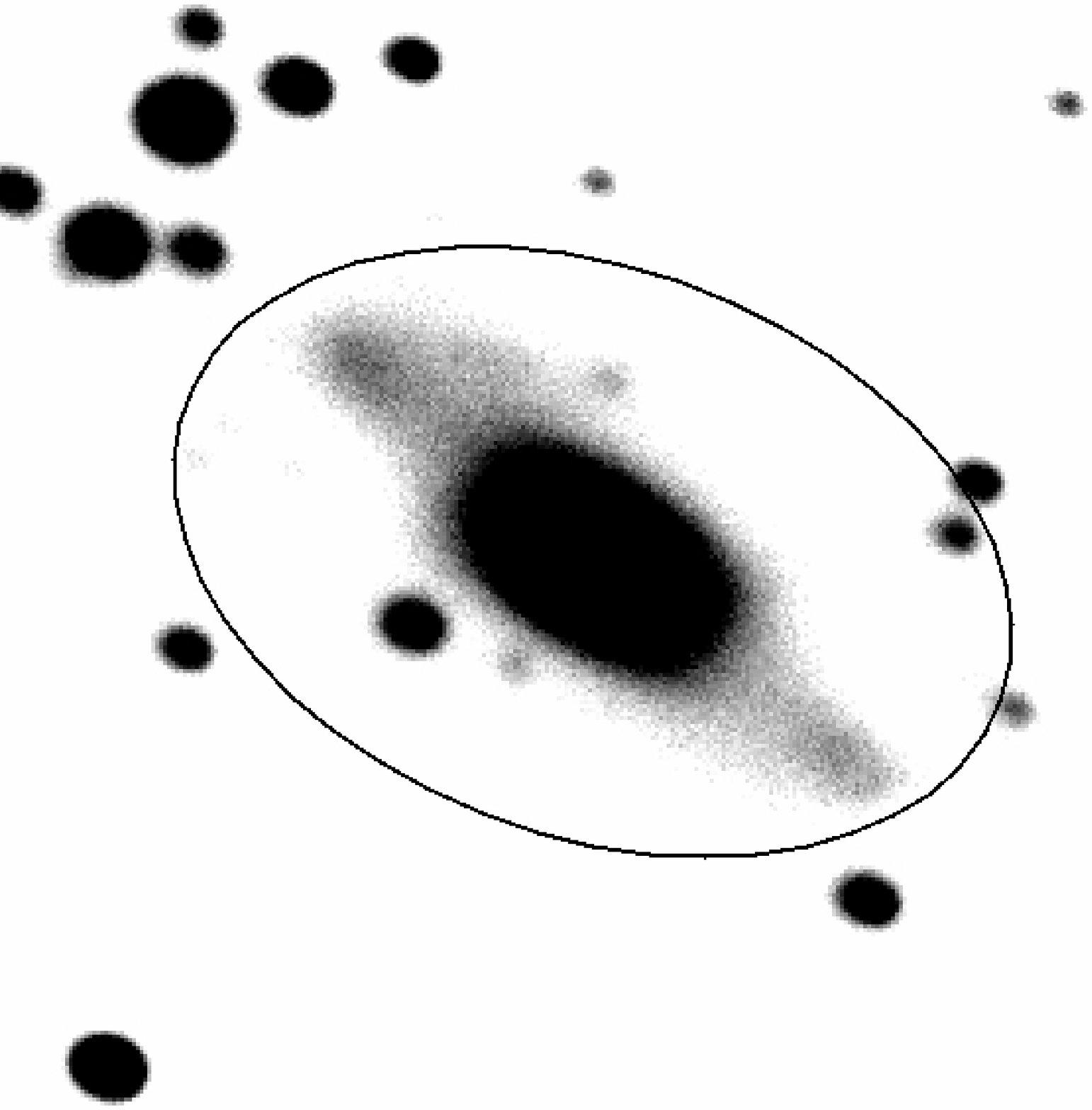}{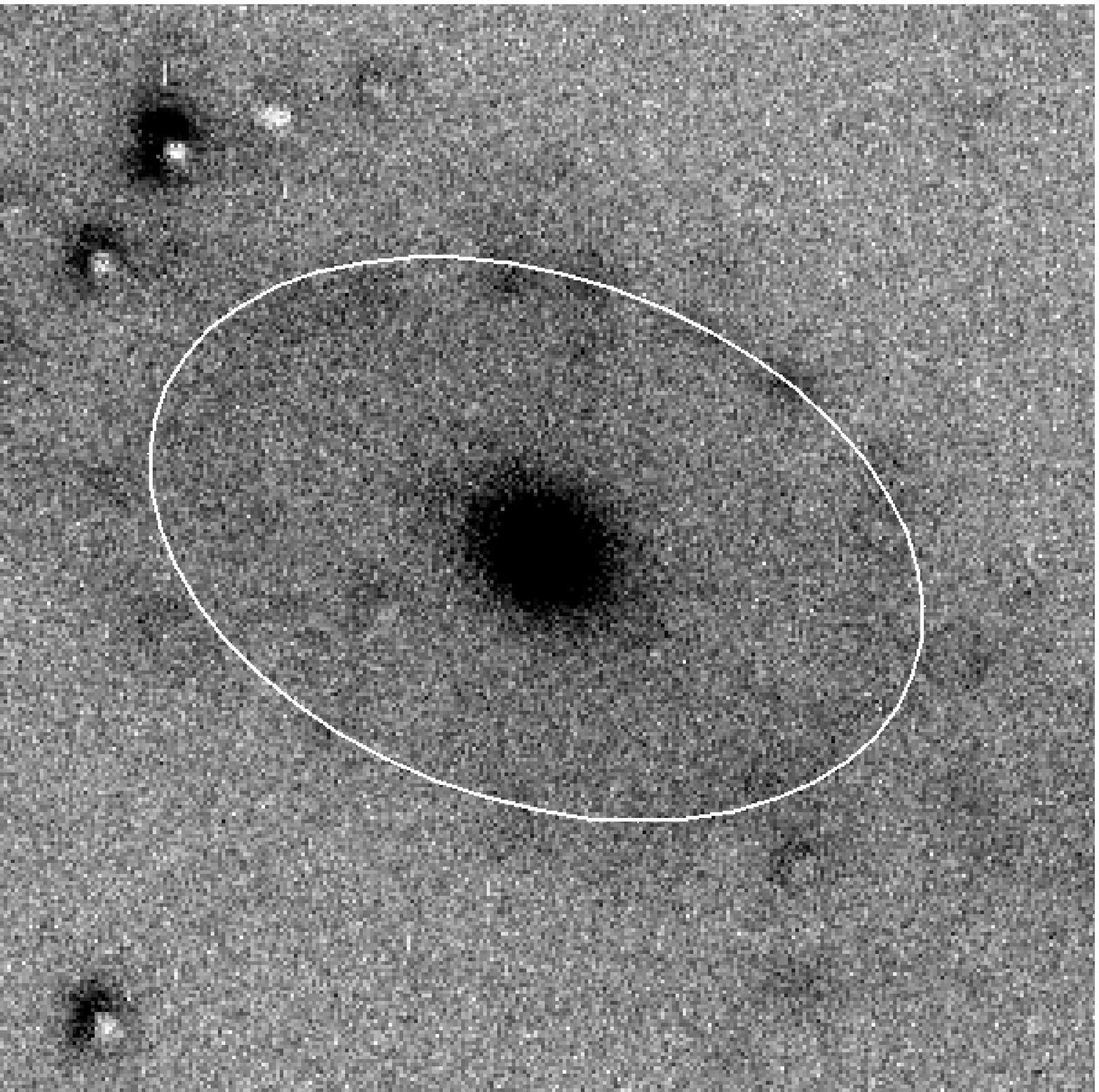}
\caption{Images of UGC~11355 with the stretch altered to show the galaxy's nuclear bar (left - R-band image)
and star forming ring (right - \Ha\ image).  In both images the ellipse shows the shape and size of the
star forming ring. The images are 1.0$^\prime$ across.  \label{fig:U11355}}
\end{figure}

Perhaps the most intriguing galaxy of our sample is UGC~11355.  This galaxy was placed in Category 2, above, as it has a 
bright nucleus and clumpy disk in the \Ha\ image.  The B and R band images of UGC~11355 show a galaxy with a simple Sbc morphology.
The \Ha\ image, though, shows a distinct star forming ring.  The ring is 
at a very different inclination from the rest of the galaxy ({\it i}=49$^\circ$ for the ring and 73$^\circ$ 
for the galaxy as a whole), and lies approximately 2.6 kpc in radius from the center of the galaxy,
measured along the major axis.  As the B and R images
show no indication of a ring morphology
this indicates unusually strong star formation in the ring.  It is also useful to note the presence of a bar
in UGC~11355 -- shown more clearly in Figure~\ref{fig:U11355}.
The fact that the inclination of the ring is 
significantly different from that of the rest of the galaxy suggests the ring a tidal effect due to an interaction,
such as a small satellite galaxy being cannibalized by UGC~11355, or the influence of 
CGCG 143-026, 14.9$^\prime$ and 68 \kms\ away.  

It is interesting to note that the \Ha\ morphology of the galaxies does not appear to correlate with
the galaxies' color profile (Figure~\ref{fig:colors}).  The galaxy with the steepest slope in the color profile
is UGC~08644 which has only a few \ion{H}{2} regions in its outer arms.  The other galaxies with steep color
profiles are UGC~00023 and UGC~8904, which have a bright knot of star formation in the nucleus 
and faint \Ha\ spread throughout their arms, and UGC~11840 and UGC~11068 both of which have no detectable \Ha.
The galaxies with the shallowest slopes similarly show no correlation between their color profiles and
morphology. This suggests that the current star formation
in these galaxies is largely independent of the past  star-formation history,
although this result should be confirmed with better, extinction-corrected, data.

\section{Star Formation}
\label{sec:StarFormation}

\begin{figure}
\plotone{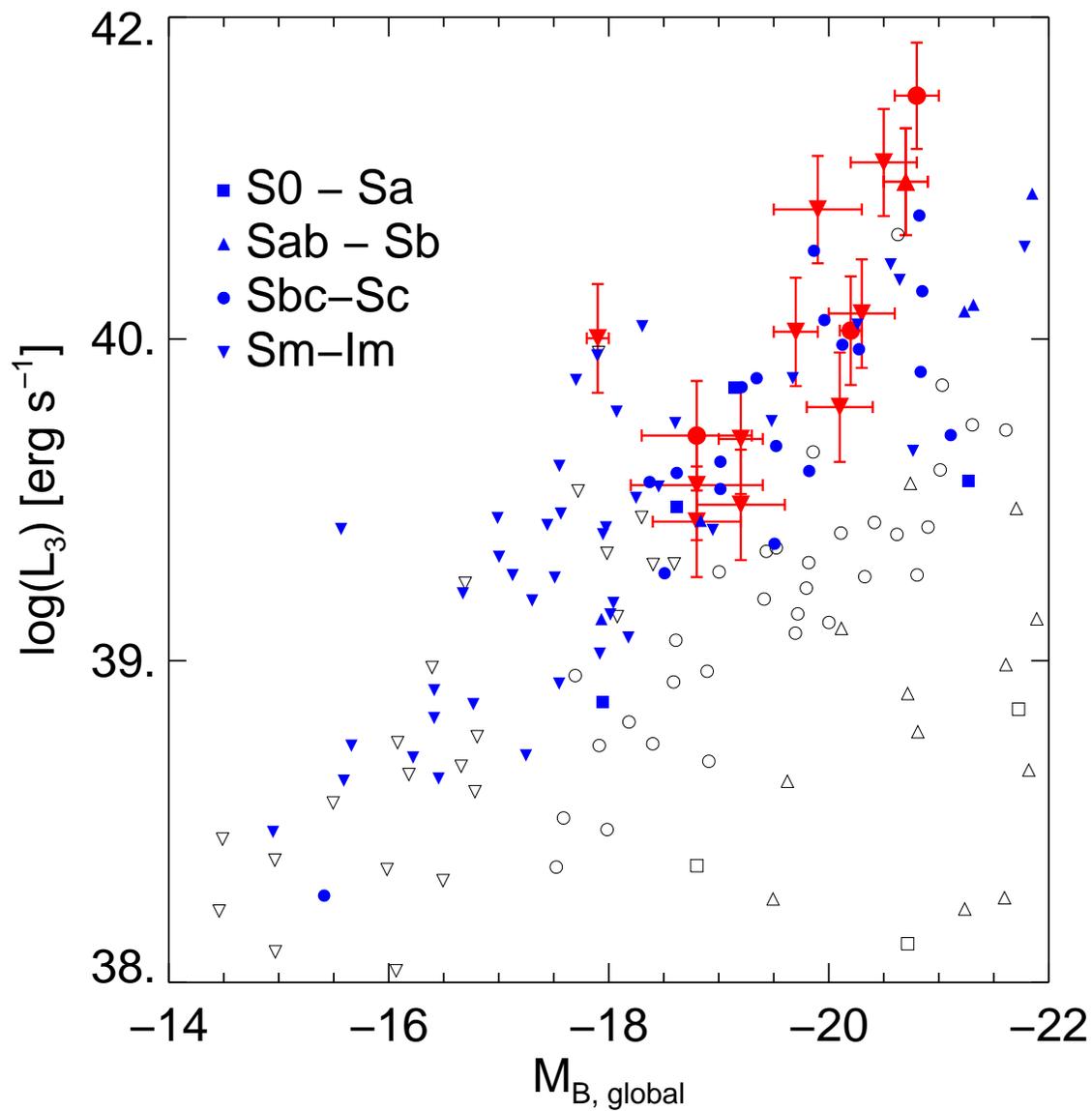}
\caption{Total B magnitude plotted against the average luminosity of the brightest three
\Ha\ regions.  (If less than three regions were found, the average of all \ion{H}{2} regions was
used.)  The filled (red) symbols are the data from our observations; the filled (blue) symbols
are from \citet{helmboldt05}; and the open (black) symbols are from \citet{kennicutt83}.
\label{fig:MB_L3}}
\end{figure}

\begin{figure}
\plottwo{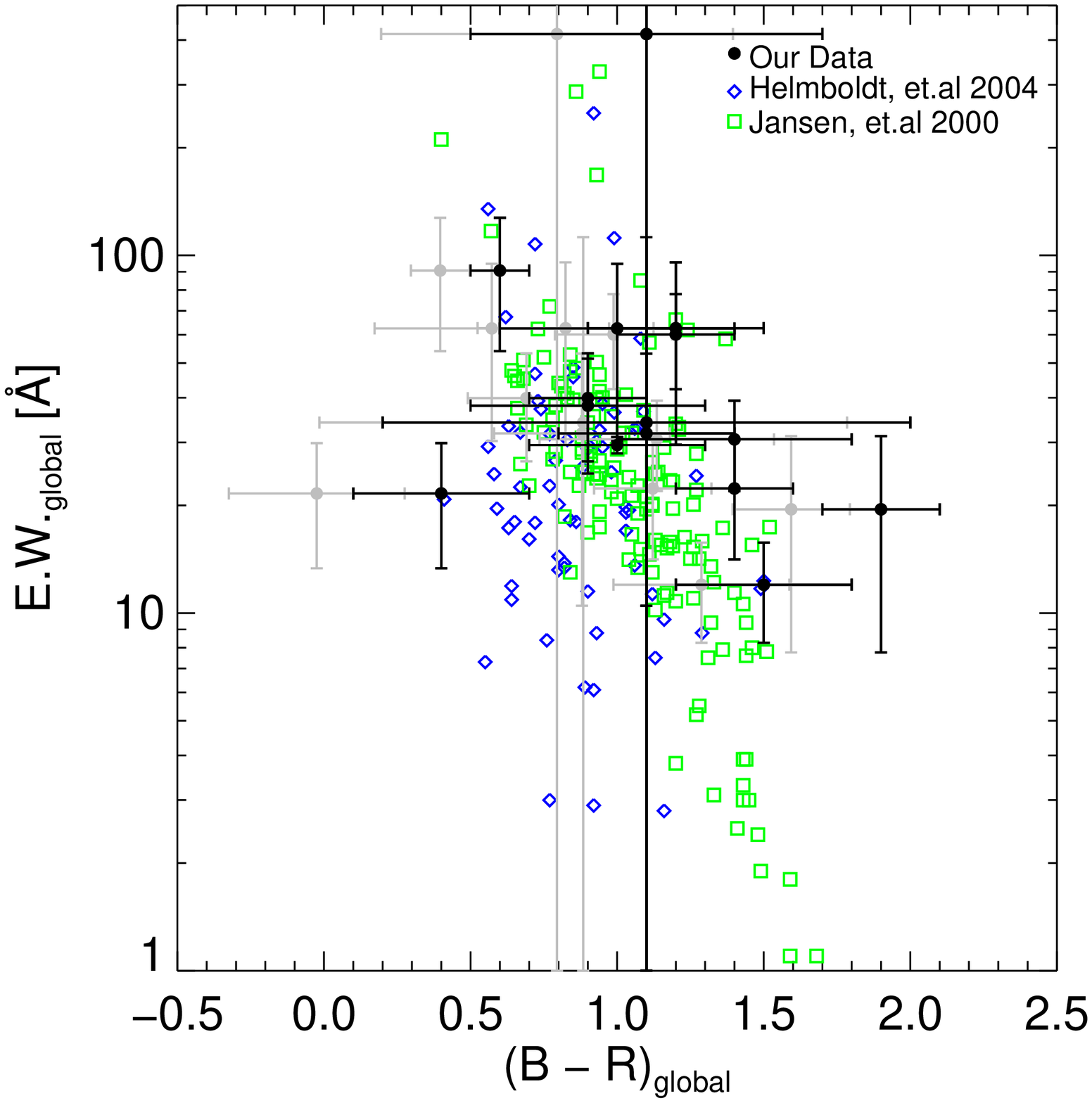}{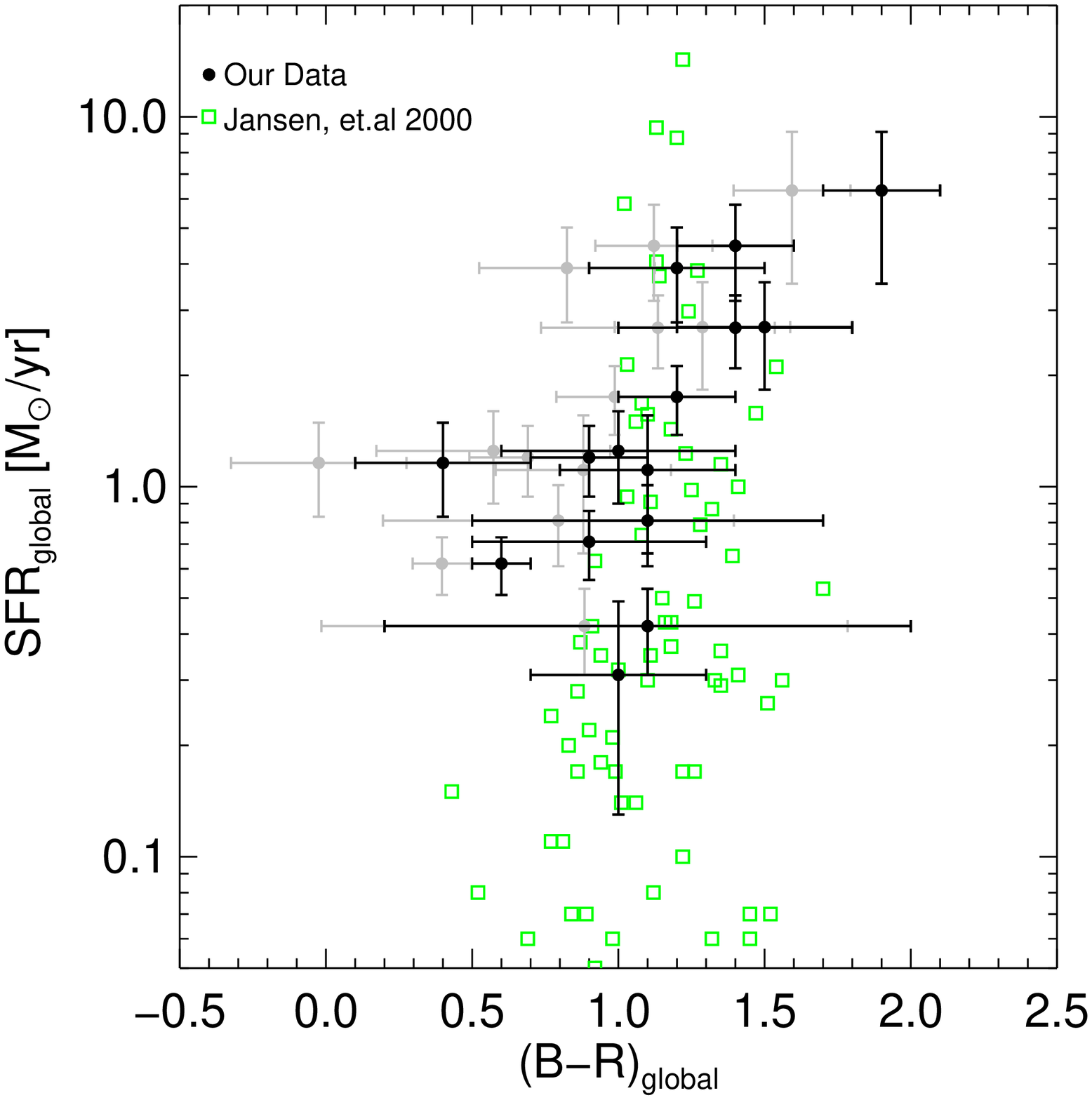}
\caption{Global color versus equivalent width (left) and star formation rate (right).
To insure any trends (or lack) remain the same, the data from this paper is shown both without
inclination correction (black) and with (gray).  Note that inclination corrections are described in
Section~\ref{sec:reduce}.
As the global SFR was not available for the \citet{helmboldt05} data, it is not
shown on the right. \label{fig:BR_EW}}
\end{figure}

\begin{figure}
\plottwo{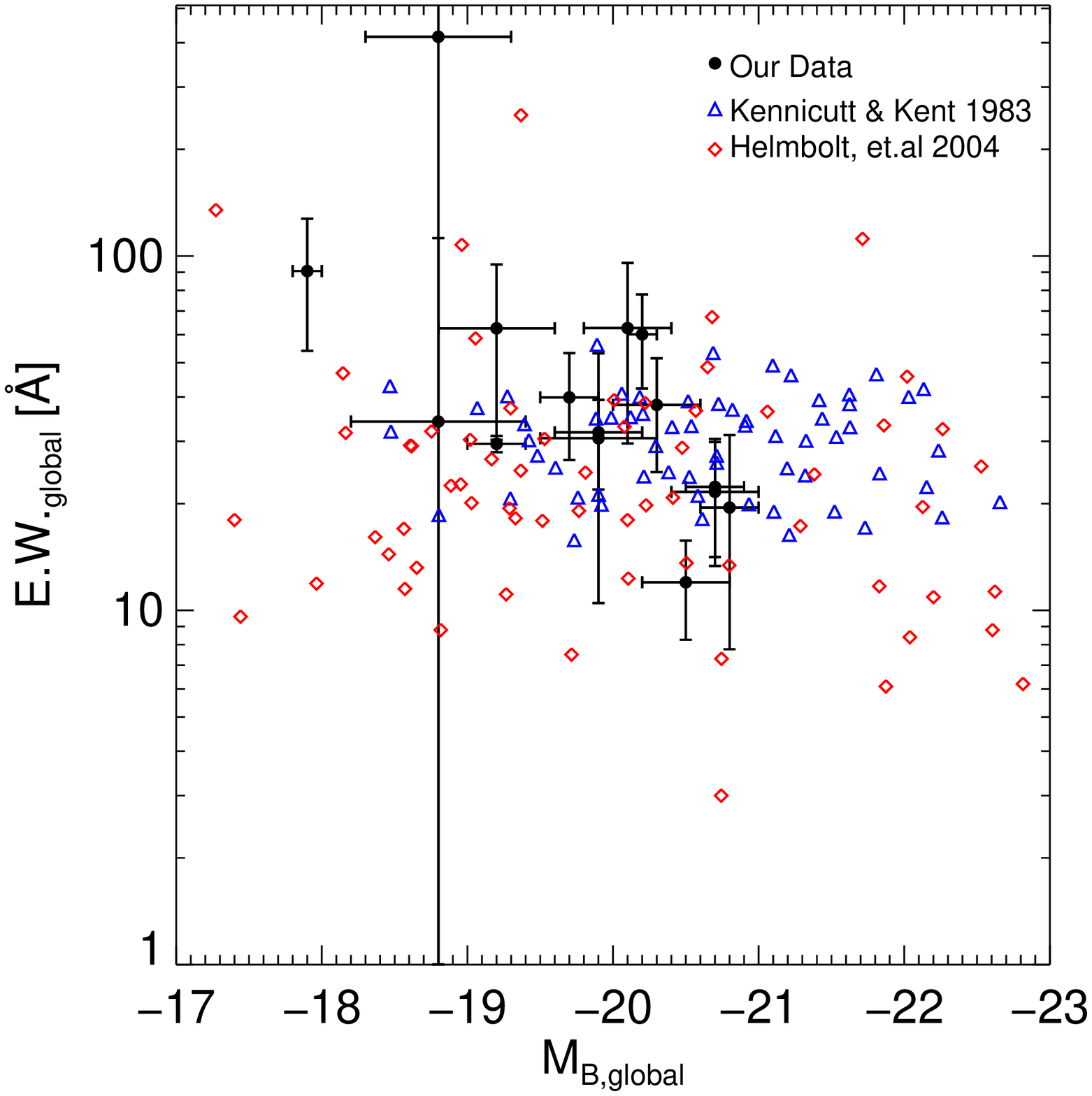}{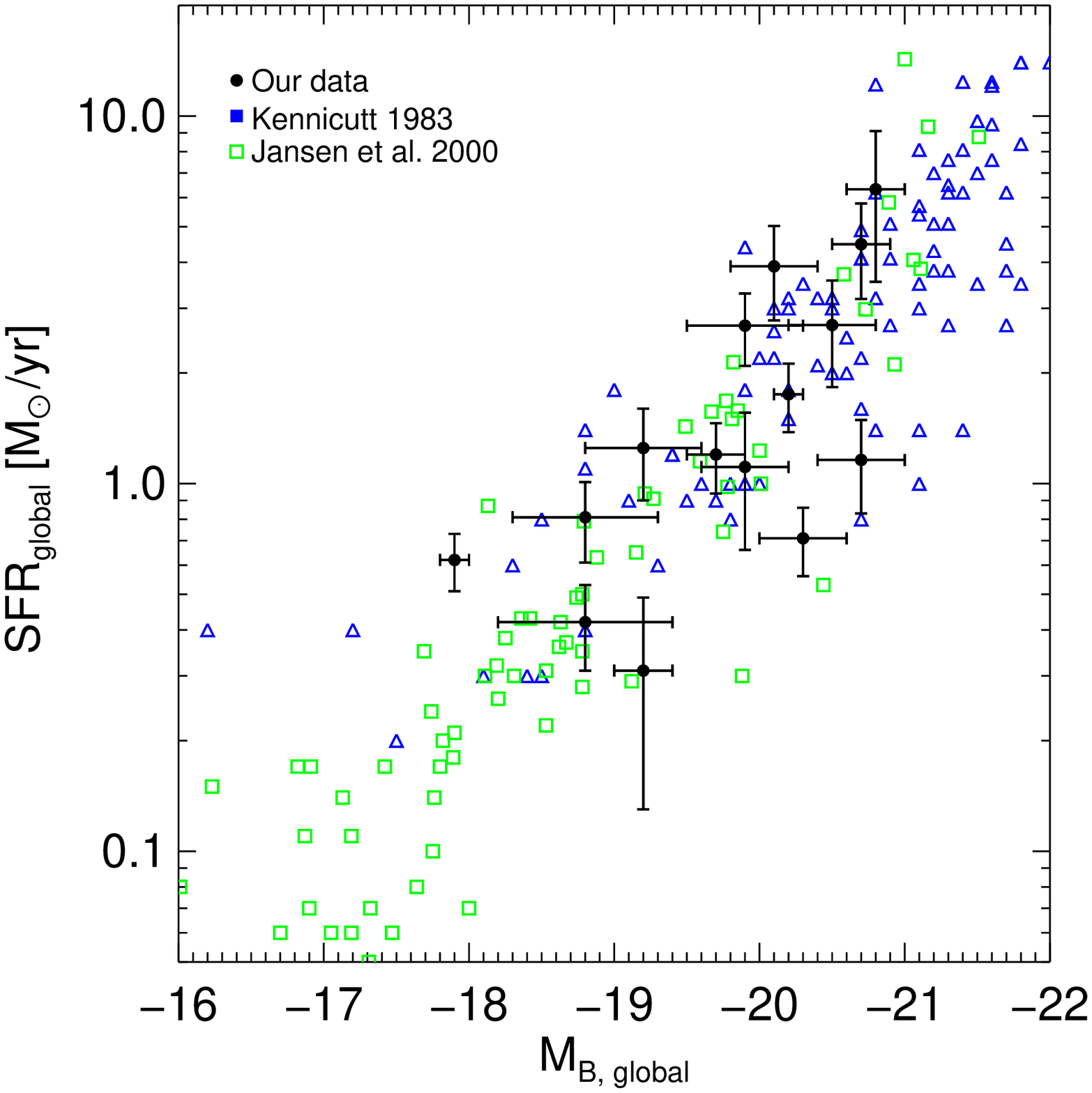}
\caption{Total B magnitude plotted against the global equivalent width (left) and star formation rate (right).
As the global SFR was not available for the \citet{helmboldt05} data, it is not
shown on the right. \label{fig:MB_EW}}
\end{figure}

\begin{figure}
\plotone{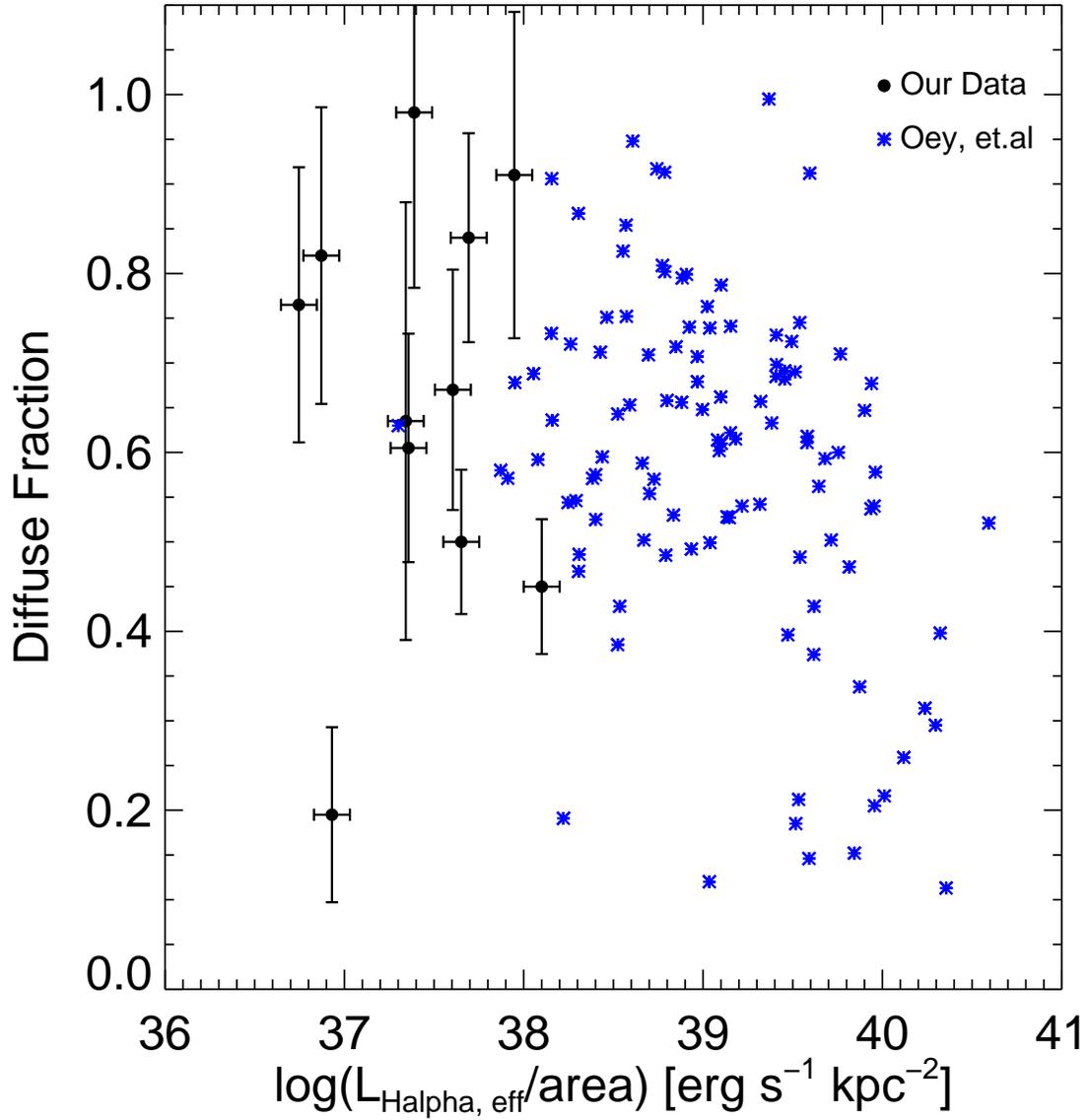}
\caption{Luminosity surface brightness (Luminosity/area) plotted against the diffuse \Ha\ fraction for our
sample and that of \cite{oey06}. \label{fig:Larea_Diffuse} }
\end{figure}

\begin{figure}
\plottwo{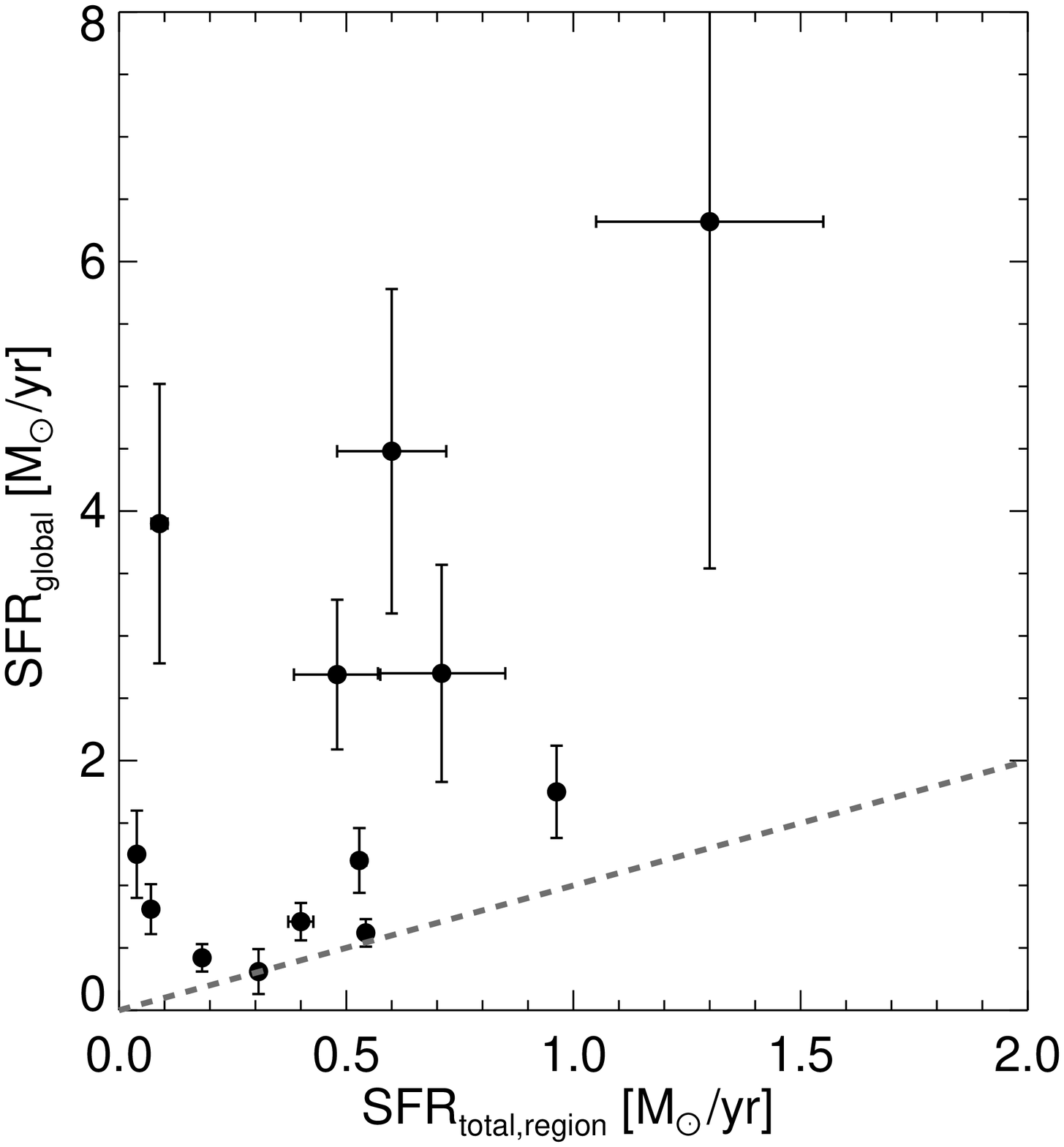}{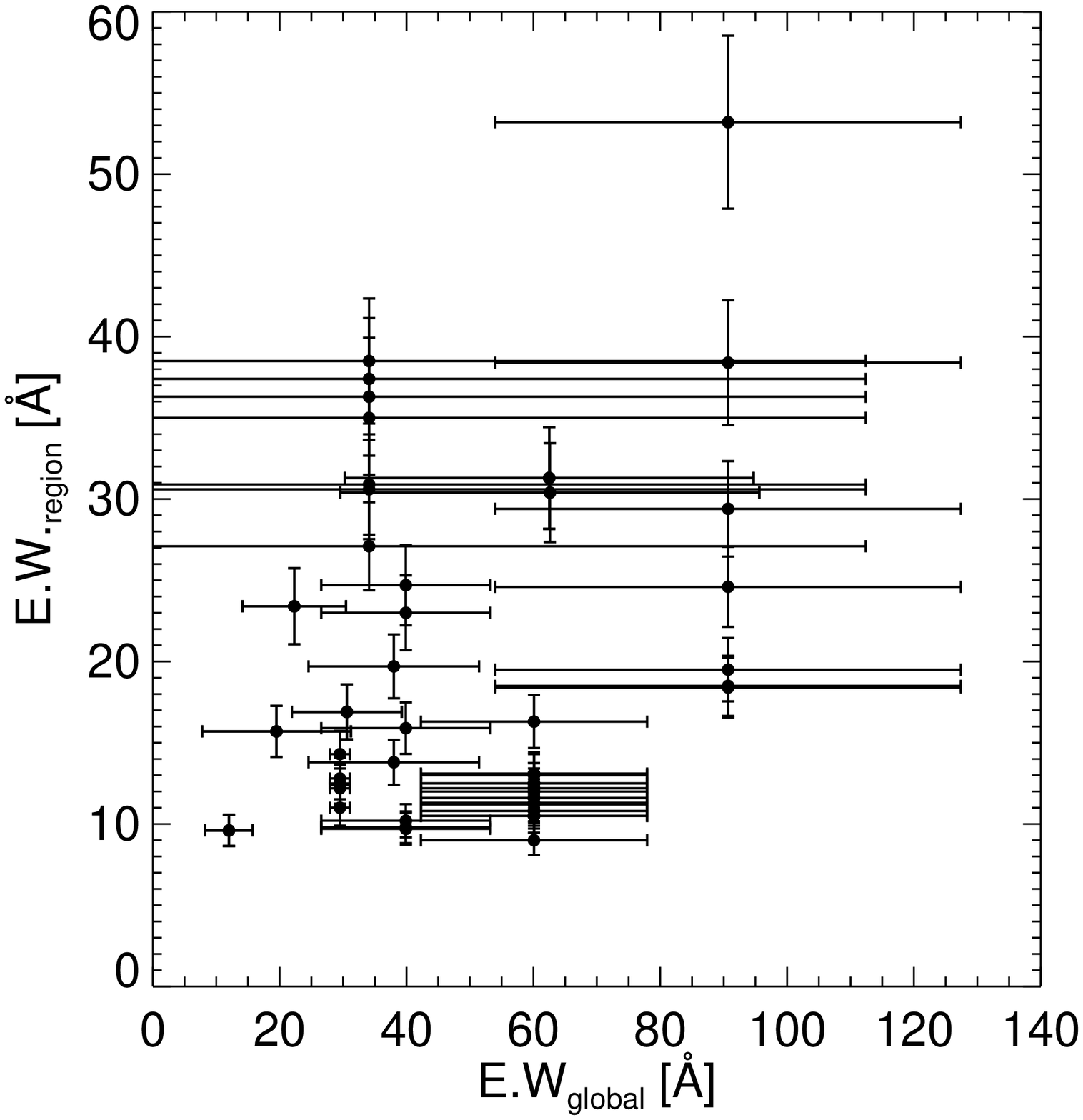}
\caption{A comparison of regional and global star formation rate and 
equivalent width for the studied galaxies.  On the
left is a plot of the global SFR against the total SFR found for the individual \ion{H}{2} regions,
with a line demarcating the point where the global and regional SFR are equal.
On the right is a plot of the global EW against the average EW for the individual \ion{H}{2} regions.
\label{fig:SFR_SFR}}
\end{figure}

\begin{figure}
\plotone{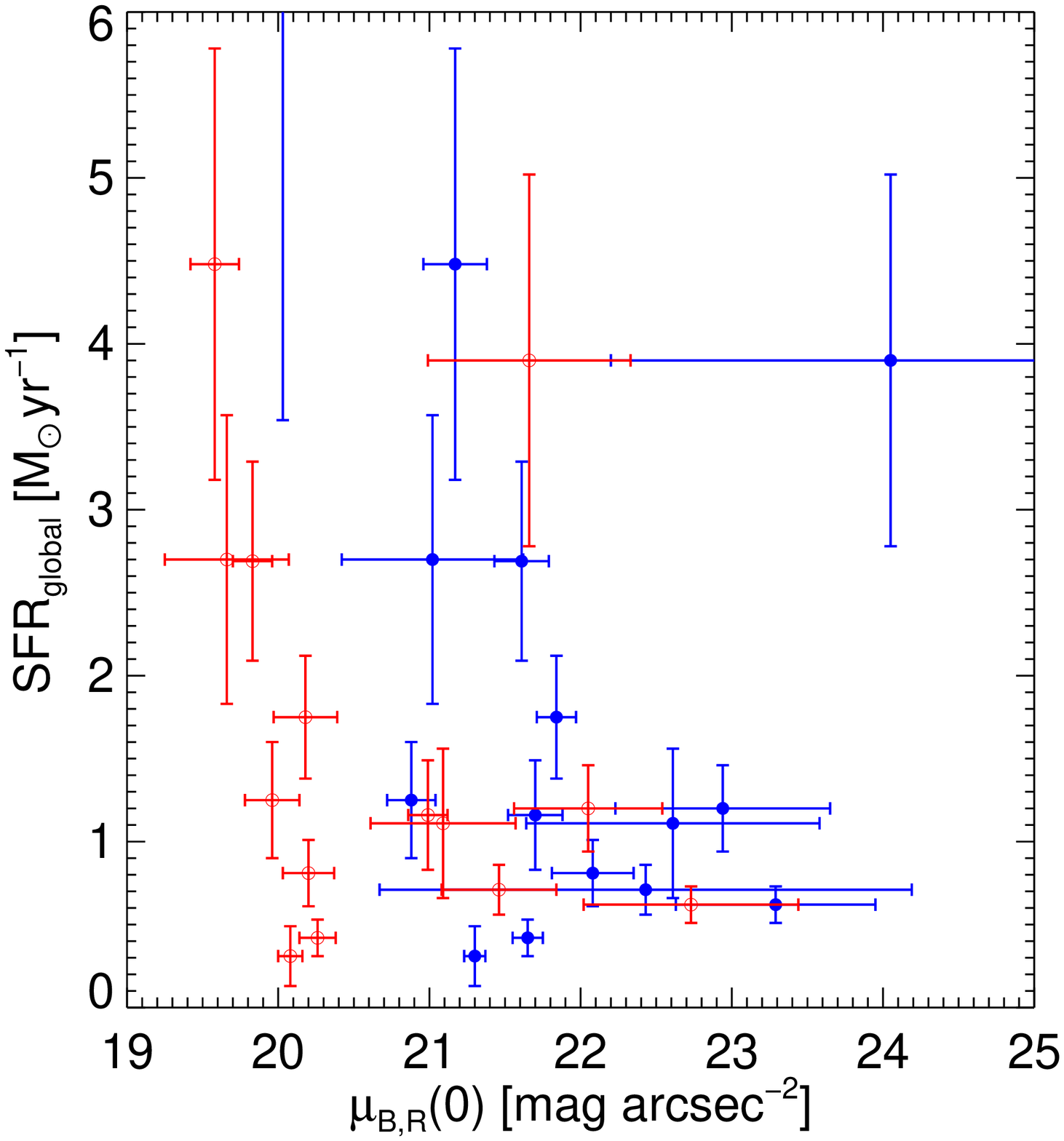}
\caption{Central surface brightness versus global star formation rate for the observed galaxies.
The (red) open circles are from the R band data, while the (blue) filled circles are for the B data.
\label{fig:mu_SFR}}
\end{figure}

\begin{figure}
\plottwo{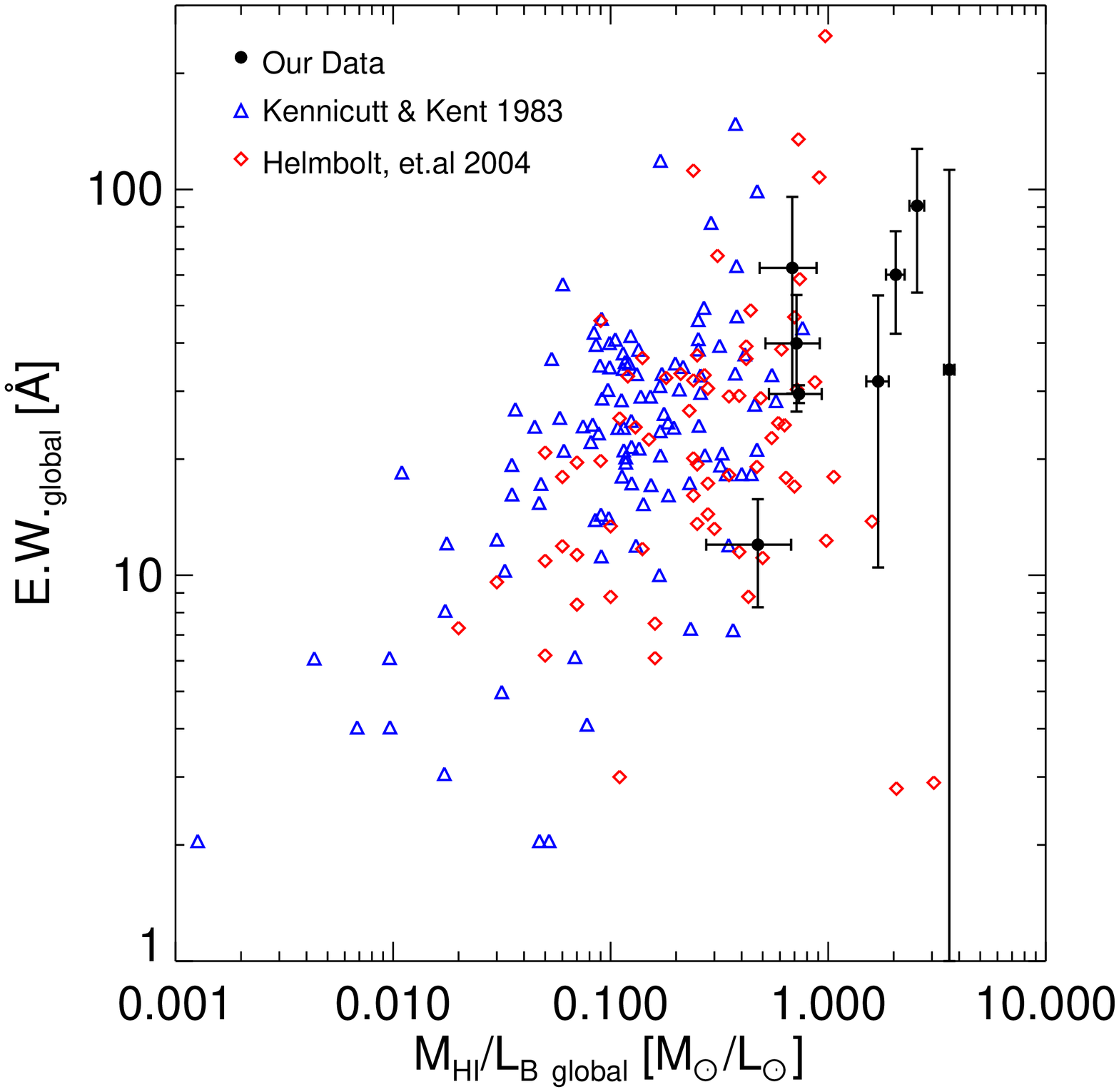}{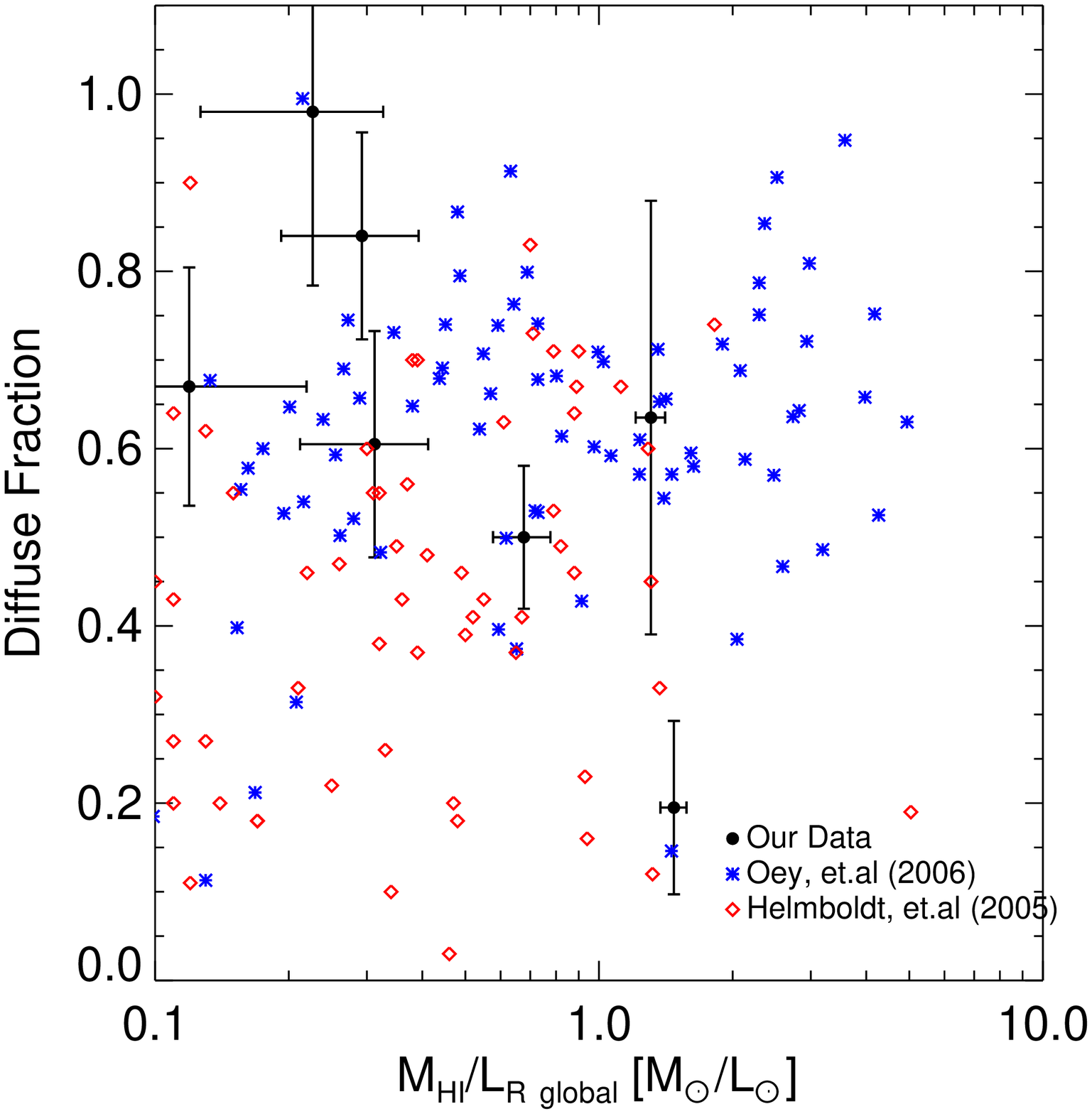}
\caption{Gas mass to luminosity ratios plotted against global equivalent widths (left) and diffuse \Ha\ fractions (right).
On the left, the (black) circles are our data, the (blue) triangles  are from \citet{kennicutt83} and the 
(red) diamonds are from \citet{helmboldt04}.  On the right, the (black) circles are again our data, while the 
(blue) asterisks are from \citet{oey06}.
\label{fig:MHILB_EW}}
\end{figure}

\begin{figure}
\plotone{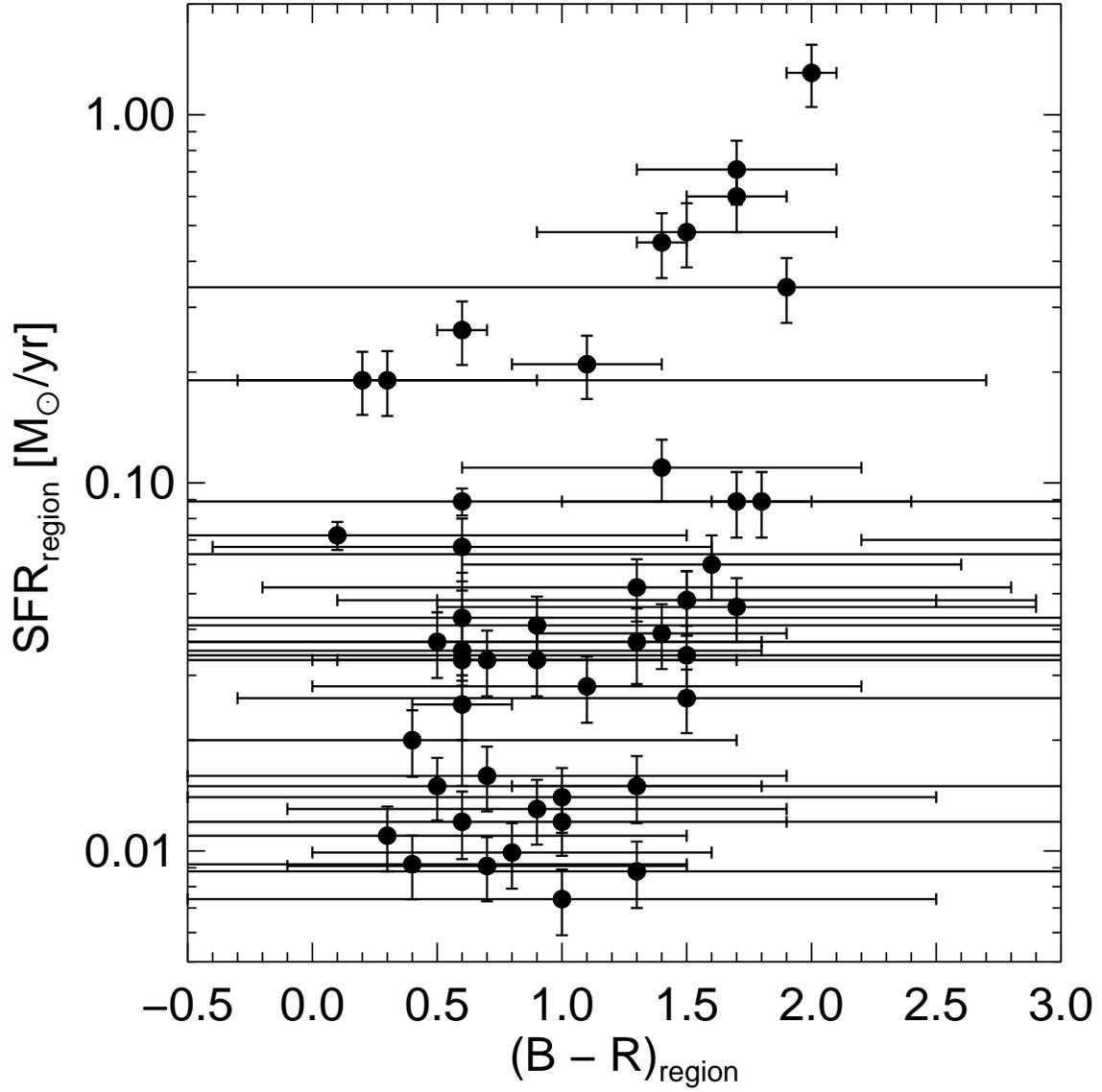}
\caption{This plot shows the regional colors versus star formation rates for the observed galaxies.
\label{fig:BR_SFR_region}}
\end{figure}

Figures~\ref{fig:MB_L3} -- \ref{fig:BR_SFR_region}  compare the properties of the \ion{H}{2} regions and emission
of our galaxy sample.  Where possible, measurements from other samples of late-type galaxies
are also shown \citep{kennicutt83, jansen00, helmboldt05, oey06}.  Examining
the figures it is clear that the overall properties of our sample are similar
to those of other late-type (Sbc-Sc) galaxies.  That is, the values for the
individual \ion{H}{2} region luminosities are similar to those
reported by \citet{helmboldt05} and  \citet{kennicutt83} (Figure~\ref{fig:MB_L3})
while the global \Ha\ equivalent width (EW) and global star formation 
rates match those seen by all three
comparison samples (Figures~\ref{fig:BR_EW}, \ref{fig:MB_EW}).  

We should note that as discussed in \S~\ref{sec:reduce} our sample suffers from having many of the \ion{H}{2} regions
blended together as a result of the distance to our galaxy samples.  As a result, it is highly likely 
that in the comparisons of the luminosities for the galaxies' individual \ion{H}{2} regions
the luminosities (Figure~\ref{fig:MB_L3}) from our sample are artificially higher then those 
in the other sample, potentially by a factor of 3 or more.  This fact does not alter the
results of this section, but it is the likely explanation for the slightly higher than average
values found for L$_3$ in Figure~\ref{fig:MB_L3}.

To examine the total amount of gas found within the \ion{H}{2} regions compared with that found in the 
diffuse \Ha\ gas, we need to determine the galaxies' \Ha\ diffuse fraction, defined
here as the ratio of \Ha\ flux not found within the defined \ion{H}{2} regions to the total \Ha\
flux found for the entire galaxy.  Examining Figures~\ref{fig:Larea_Diffuse} and \ref{fig:SFR_SFR}, as well
as Tables~\ref{tab:global2} and \ref{tab:regions},
reveals an interesting fact -- while the global
SFR for these galaxies is fairly typical (0.3 -- 5 \Msol/yr), the combined SFR from the
galaxies' \ion{H}{2} regions is a factor of 2 -- 10 smaller.  That is, on average the majority of the
\Ha\ emission and thus the majority of the star formation in the observed galaxies
comes not from the bright knots of star formation but instead from the galaxies' diffuse \Ha\
gas.  This is in contrast to the behavior seen from typical HSB galaxies, as evidenced by the 
data of \citet{oey06} in Figure~\ref{fig:Larea_Diffuse}.
We note that blending and angular resolution effects appear to be
relatively unimportant in estimating the fraction of diffuse \Ha\
emission.  \citet{oey06} demonstrate this by showing no systematic
changes in measured diffuse fractions as a function of distance up to
almost 80 Mpc, and inclination angle, for their sample of 100+ SINGG survey
galaxies.

While at first glance the higher diffuse \Ha\ fractions found for these galaxies seems surprising, 
recent GALEX results of the outer edges of M83, a region whose environment closely resembles that of the disks of
massive LSB galaxies,  also show considerable star formation outside the \ion{H}{2}
regions in that part of the galaxy \citep{thilker05}.  Similarly, \citet{helmboldt05} found a slight trend with lower
surface brightness galaxies having higher diffuse fractions than their higher surface brightness counterparts.

The fact that these galaxies have higher \Ha\ diffuse gas fractions raises an interesting question.  Typically
diffuse gas is believed to be ionized by OB stars lying within density-bounded \ion{H}{2} regions.  The problem 
of transporting the ionizing photons from these regions to the diffuse gas is extreme in these cases, as 
there would need to be a very large number of density-bound \ion{H}{2} regions leaking ionizing photons 
to ionize the quantity of diffuse gas seen here.  \citep[See the more detailed discussion in ][ which also discusses
shock heating from stellar winds and SNe as ionization sources.]{hoope01}
An alternative suggestion is that field OB stars are also ionizing the diffuse gas, as was suggested by 
\citet{hoope01}.  This would imply a different stellar population within and without the \ion{H}{2} regions, as it 
would likely be the later OB types (B0--O9) which either escape the \ion{H}{2} regions or survive the regions' 
destruction.  We note \citet{oey04} predict a modest increase in the fraction of field
massive stars in galaxies with the lowest absolute star-formation rates.  Scheduled GALEX observations of a subset of our observed galaxies may shed light on the underlying 
stellar population in the galaxies' diffuse stellar disks.

Finally, it is elucidating to look for any trends between the global and regional properties of the 
galaxies and their SFR and \Ha\ content.  Figure~\ref{fig:mu_SFR} plots the galaxies' central surface brightness
(in both B and R) against the galaxies' total star formation rate.  While the error bars make defining any trend
difficult, there certainly appears to be a decrease in the global SFR with decreasing central surface brightness,
similar to the trends seen in other studies \citep[e.g.][]{vandenhock00,gerritsen99}. 
Figure~\ref{fig:MHILB_EW} shows the galaxies' gas-to-luminosity
ratios plotted against both their global equivalent width and diffuse \Ha\ fraction.  In both cases, no trend
can be seen with our data, although the small number of points available make any diagnosis difficult.  
Combined with the other datasets, though, we can see a general trend toward higher equivalent widths with
increasing M$_{HI}$/L$_B$, but surprisingly no trend between gas fraction and the galaxies' diffuse \Ha\ fraction is 
visible.  This lack of correlation is also seen by \citet{oey06}. 
The last trend which can be seen is a rough correlation between the galaxies' global color and 
star formation rate (Figure~\ref{fig:BR_EW}), with redder galaxies having higher SFR, a fact
which may be a reddening effect. 
The individual \Ha\ regions, however, show no such trend (Figure~\ref{fig:BR_SFR_region}).

\section{Conclusion}

The sample of 19 galaxies observed for this project were chosen to be large galaxies with low surface
brightness disks.  The surface brightness measurements for this sample were obtained originally through the 
UGC measurements through determining the galaxies' average surface brightness within the $\mu$=25 \mss\ isophote.
The relation employed to determine the galaxies' average surface brightness (Equation~\ref{eqn:bothun}) has
shown itself it be a good predictor of a galaxy's central surface brightness.  But for a wide variety of reasons
the UGC measurements were not sufficient to insure the galaxies contained within this catalog have true 
LSB disks, underscoring the difficulty in designing targeted searches for large LSB galaxies.
 
Nonetheless, the sample of galaxies observed for this project have lower surface brightnesses than is found for a
typical sample of large high surface brightness galaxies.  In most other aspects the galaxies appear fairly 
`normal', with colors typically B$-$R=0.3$-$0.9, morphological types ranging from Sb -- Irr, and color gradients which 
typically grow bluer toward the outer radius.  However, the galaxies have both higher gas mass-to-luminosity
fractions and diffuse \Ha\ fractions than is found in higher surface brightness samples. 
This raises two questions.  First, if the SFR for these galaxies has been similar to their
higher surface brightness counterparts through the galaxies' life, why do the lower surface brightness
galaxies have higher gas mass-to-luminosity ratios?  Second, why do these galaxies have a higher
fraction of ionizing photons outside the density-bounded \ion{H}{2} regions then their higher surface brightness
counterparts?

The answer to the first question posed above likely comes from the difference between the studied galaxies'
current and historical SFR.  As these galaxies have on average and lower metallicities 
\citep{denaray04, gerritsen99} than their higher surface brightness counterparts, it is likely that 
the galaxies' SFR has not remained constant throughout the their lifetimes.  Indeed the simplest explanation for the 
current similar SFRs and higher gas mass-to-luminosity ratios for the studied galaxies than for their higher 
surface brightness counterparts is that the galaxies' past SFR was significantly different than is currently seen. 
In fact, the measured properties would be expected if the galaxies in this study have episodic
star formation histories, with significant time (1-3 Gyr) lapsing between
SF bursts, as has been conjectured for LSB galaxies in the past \citep[e.g.][]{gerritsen99}.  
Such a star formation history would help promote significant changes in the galaxies' mean surface
brightness and allow an individual large disk galaxy to appear as either (a) a relatively
normal Hubble sequence spiral, (b) a large, lower surface brightness disk, or (c) perhaps
even a lower surface brightness disk if the time between episodes is sufficiently large,
depending on the elapsed time since the last SF burst.  
The final answer to this may
be found when an answer to the second question, determining why the diffuse fractions for
the studied galaxies is higher than for similar HSB galaxies, is also found.  Irregardless,
what is clear is that the studied sample shows a clear bridge between the known properties
of high surface brightness galaxies and the more poorly understood properties of their very low surface 
brightness counterparts, such as Malin 1.

\acknowledgements
Thanks to Joe Helmboldt for his help in getting the HIIphot program running with
LSB galaxies and to Rene Walterbos for his loan of the \Ha\ filters.
MSO acknowledges support from the National Science Foundation, grant AST-0448893.

\clearpage
\thispagestyle{empty}
%
%
\begin{deluxetable}{ccccccccccccccccc}
\rotate
\tabletypesize{\scriptsize}
\tablewidth{0pt}
\tablecaption{Global Properties of Galaxies -- B \& R Measurements\label{tab:global}}
\tablehead{
& & & & &
\multicolumn{4}{c}{B} & & \multicolumn{4}{c}{R}\\
\cline{6-9} \cline{11-14}\\
\colhead{Galaxy}& \colhead{RA\tablenotemark{a}} &\colhead{Dec\tablenotemark{a}} & \colhead{Vel\tablenotemark{a}} & \colhead{Type\tablenotemark{a}} &
\colhead{m\tablenotemark{b}} & \colhead{M\tablenotemark{b}} &\colhead{D$_{25}$\tablenotemark{c}} &\colhead{$\langle \mu \rangle$\tablenotemark{d}}&
&\colhead{m\tablenotemark{b}} & \colhead{M\tablenotemark{b}} &\colhead{D$_{25}$\tablenotemark{c}} &\colhead{$\langle \mu \rangle$\tablenotemark{d}}&
\colhead{r\tablenotemark{b}} & \colhead {B$-$R\tablenotemark{b}} & \colhead {$i$\tablenotemark{e}}\\
\\
&\colhead{$\left[J2000\right]$} & \colhead{$\left[J2000\right]$}& \colhead{$\left[km\;s^{-1}\right]$}
& &\colhead{$\left[mag\right]$} & \colhead{$\left[Mag\right]$} &\colhead{$\left[^{\prime\prime}\right]$}& 
\colhead{$\left[mag/\prime\prime^{2}\right]$}  &
&\colhead{$\left[mag\right]$} & \colhead{$\left[Mag\right]$} &\colhead{$\left[^{\prime\prime}\right]$}&
\colhead{$\left[mag/\prime\prime^{2}\right]$} & 
\colhead{$\left[^{\prime\prime}\right]$} & & \colhead{$\left[{^\circ}\right]$}
}
\startdata
  UGC  00023 &  00 04 13.0 & 10 47 25  &  7787  &  3  &   14.4 (0.1) & -20.7 (0.1) &   71  &  23.4  && 13.0 (0.1) & -22.1 (0.1) &  86.7 &  22.4  &  40   &  1.4  (0.2) & 52\\
  UGC  00189 &  00 19 57.5 & 15 05 32  &  7649  &  7  &   15.0 (0.2) & -20.1 (0.2) &   84  &  24.4  && 13.8 (0.1) & -21.3 (0.1) & 116.1 &  23.8  &  57   &  1.2  (0.2) & 67\\
  UGC  01362 &  01 52 50.7 & 14 45 52  &  7918  & 8.8 &   16.9 (0.2) & -18.2 (0.2) &   30  &  24.3  && 15.6 (0.1) & -19.5 (0.1) &  42.1 &  23.5  &  23   &  1.3  (0.3) &  0\\ 
  UGC  02299 &  02 49 07.8 & 11 07 09  & 10253  &  8  &   15.4 (0.2) & -20.3 (0.2) &   59  &  24.0  && 14.5 (0.4) & -21.2 (0.4) &  65.1 &  23.3  &  33   &  0.9  (0.4) & 32\\ 
  UGC  02588 &  03 12 26.5 & 14 24 27  & 10093  & 9.9 &   15.8 (0.2) & -19.9 (0.2) &   39  &  23.6  && 14.7 (0.1) & -20.9 (0.1) &  50.1 &  23.0  &  28   &  1.1  (0.2) & 28\\ 
  UGC  02796 &  03 36 52.5 & 13 24 24  &  9076  &  4  &   14.8 (0.2) & -20.6 (0.2) &   66  &  23.6  && 13.3 (0.1) & -22.1 (0.1) &  94.1 &  22.7  &  28   &  1.5  (0.2) & 57\\ 
  UGC  03119 &  04 39 07.7 & 11 31 50  &  7851  &  4  &   14.3 (0.2) & -20.8 (0.2) &   71  &  23.7  && 12.4 (0.1) & -22.7 (0.1) &\tablenotemark{\ddag}  &  24.1  &  40   &  1.9  (0.2) & 72\\ 
  UGC  03308 &  05 26 01.8 & 08 57 25  &  8517  &  6  &   14.3 (0.3) & -21.0 (0.3) &   89  &  23.8  && 14.0 (0.2) & -21.3 (0.2) &  88.1 &  23.5  &  48   &  0.3  (0.4) & 28\\ 
  UGC  07598 &  12 28 30.9 & 32 32 52  &  9041  & 5.9 &   15.3 (0.1) & -20.1 (0.1) &   46  &  23.5  && 14.8 (0.1) & -20.6 (0.1) &  66.8 &  22.8  &  33   &  0.5  (0.2) & 22\\ 
  UGC  08311 &  13 13 50.8 & 23 15 16  &  3451  & 4.1 &   15.5 (0.1) & -17.9 (0.1) &   50  &  23.8  && 14.8 (0.1) & -18.5 (0.1) &  62.0 &  23.5  &  33   &  0.7  (0.2) & 26\\ 
  UGC  08644 &  13 40 01.4 & 07 22 00  &  6983  &  8  &   16.1 (0.2) & -18.8 (0.2) &   43  &  24.2  && 15.3 (0.2) & -19.6 (0.2) &  49.4 &  23.5  &  28   &  0.8  (0.3) & 30\\ 
  UGC  08904 &  13 58 51.1 & 26 06 24  &  9773  & 3.6 &   15.9 (0.1) & -19.7 (0.1) &   43  &  23.8  && 14.9 (0.1) & -20.7 (0.1) &  55.5 &  23.4  &  33   &  1.0  (0.2) & 48\\ 
  UGC  10894 &  17 33 03.8 & 27 34 29  &  6890  &  4  &   16.0 (0.3) & -18.8 (0.3) &   48  &  24.1  && 14.9 (0.3) & -19.9 (0.3) &\tablenotemark{\ddag}  &  23.0  &  28   &  1.1  (0.4) & 57\\ 
  UGC  11068 &  17 58 05.0 & 28 14 38  &  4127  & 3.2 &   15.0 (0.2) & -18.7 (0.2) &   64  &  24.0  && 13.8 (0.1) & -19.9 (0.1) &  89.4 &  23.4  &  57   &  1.2  (0.2) &  0\\ 
  UGC  11355 &  18 47 57.0 & 22 56 33  &  4360  & 3.5 &   13.9 (0.3) & -19.9 (0.3) &  173  &  24.8  && 12.5 (0.1) & -21.3 (0.1) &\tablenotemark{\ddag}  &  23.7  &  82   &  1.4  (0.3) & 73\\ 
  UGC  11396 &  19 03 49.5 & 24 21 28  &  4441  & 3.5 &   14.8 (0.3) & -19.1 (0.3) &  114  &  24.4  && 13.8 (0.2) & -20.1 (0.2) &  78.7 &  23.0  &  33   &  1.0  (0.4) & 59\\ 
  UGC  11617 &  20 43 39.3 & 14 17 52  &  5119  & 6.1 &   14.9 (0.2) & -19.2 (0.2) &   75  &  24.1  && 13.9 (0.2) & -20.3 (0.2) &  86.1 &  23.3  &  40   &  1.1  (0.3) & 58\\ 
  UGC  11840 &  21 53 18.0 & 04 14 50  &  7986  &  4  &   16.3 (0.1) & -18.9 (0.1) &   24  &  22.9  && 15.3 (0.1) & -19.9 (0.1) &  23.3 &  21.9  &  11   &  1.0  (0.2) & 40\\ 
  UGC  12021 &  22 24 11.6 & 06 00 12  &  4472  &  3  &   14.7 (0.2) & -19.2 (0.2) &  113  &  24.7  && 13.6 (0.1) & -20.2 (0.1) & 130.4 &  23.9  &  57   &  1.1  (0.2) & 63\\ 
\enddata
\tablecomments{Errors are given in parenthesis.}
\tablenotetext{a}{RA, Dec, velocity and type information obtained from NED, the NASA Extragalacitc Database. Galaxy types are defined in \citet{rc3}.}
\tablenotetext{b}{Magnitudes and colors were obtained at the maximum usable radius, r.  Corrections applied
and error estimates are described in Section~\ref{sec:reduce}.}
\tablenotetext{c}{D$_{25}$ are the diameters for the 25 mag arcsec$^{-2}$ isophotes.}
\tablenotetext{d}{Average surface brightness, as defined by Equation~\ref{eqn:bothun} in Section~\ref{sec:sb}.}
\tablenotetext{e}{Inclinations are simply the major to minor axis ratio of the galaxies, found through isophote fitting.}
\tablenotetext{\ddag}{Isophotes did not reach 25 mag arcsec$^{-2}$.  }
\end{deluxetable}


\clearpage
%
%
\begin{deluxetable}{ccccc}
\tabletypesize{\scriptsize}
\tablewidth{0pt}
\tablecaption{Global Properties of Galaxies -- H$\alpha$ \label{tab:global2}}
\tablehead{
\colhead{Galaxy}& \colhead{H$\alpha$ flux $\times 10^{-13}$\tablenotemark{a}}
& \colhead{ EW\tablenotemark{b} }& \colhead{SFR\tablenotemark{c}} &
\colhead{r\tablenotemark{d}} \\
&\colhead{$\left[erg \; cm^{-2} \; s^{-1}\right]$}
&\colhead{$\rm \left[\AA\right]$} &\colhead{$\rm \left[{M_\odot
yr^{-1}}\right]$} & \colhead{$\rm \left[^{\prime\prime}\right]$}
}
\startdata
UGC 00023 &	5    (1)    &	22  (8)      &	4   (1)   &	28  \\
UGC 00189 &	4    (1)    &	63  (33)     &	4   (1)	  &	23  \\
UGC 01362 &  \nodata	&   \nodata	       &  \nodata	  & \nodata  \\
UGC 02299 &	0.41 (0.09) &	38  (13)     &	0.7 (0.2) &	6  \\
UGC 02588 &	0.7  (0.3)  &	32  (21)     &	1.1 (0.5) &	13  \\
UGC 02796 &	2.1  (0.7)  &	12  (4)      &	2.7 (0.9) &	13  \\
UGC 03119 &	7    (4)    &	20  (12)     &	6 (6)	  &	28  \\
UGC 03308 &	1.0  (0.3)  &	22  (8)      &	1.1 (0.3) &	11  \\
UGC 07598 &	1.3  (0.3)  &	60. (18)     &	1.8 (0.4) &	19  \\
UGC 08311 &	3.1  (0.6)  &	91  (37)     &	0.6 (0.1) &	19  \\
UGC 08644 &     \nodata	&   \nodata	       &  \nodata	  & \nodata  \\
UGC 08904 &	0.8  (0.2)  &	40  (13)     &	1.1 (0.3) &	16  \\
UGC 10894 &	0.4  (1)    &	34  (78)     &	0.4 (0.1) &	9  \\
UGC 11068 & \nodata	&   \nodata	       &  \nodata	  & \nodata  \\
UGC 11355 &	8    (2)    &	30  (9)      &	2.7 (0.6) &	28  \\
UGC 11396 &	2.4  (0.8)  &	400 (2000)   &	0.8 (0.2) &	16  \\
UGC 11617 &	3.1  (0.7)  &	30  (2)	 &	0.3 (0.2) &	19  \\
UGC 11840 & \nodata	&   \nodata	       &  \nodata	  & \nodata  \\
UGC 12021 &	4    (1)    &	63  (32)     &	1.3 (0.4) &	23  \\
UGC 12289 & \nodata     &   \nodata	       &  \nodata	  & \nodata  \\
\enddata
\tablecomments{Derivation of quantities are described in
Section~\ref{sec:reduce}. Errors are given in parenthesis.}
\tablenotetext{a}{Total \Ha\ flux found within the radius centered on the
(optical) center of the galaxy and extending
to the radius given in the last column. Errors were determined in the same
manners as for magnitudes, and are given in
Section~3.}
\tablenotetext{b}{The equivalent width was calculated simply as the ratio of
the total \Ha\ flux to
total \Ha-subtracted continuum flux.}
\tablenotetext{c}{SFR = ${{L_{H\alpha}}\over{1.26\times 10^{41} erg s^{-1}}}$;
from \citet{kennicutt94}.}
\tablenotetext{d}{Radius at which the isophotal signal-to-noise went below
1$\sigma$. }
\end{deluxetable}


\begin{deluxetable}{rccccccccc}
\tabletypesize{\scriptsize}
\tablewidth{0pt}
\tablecaption{Fitted Galaxy Properties\label{tab:fitted}}
\tablehead{
&&&
\multicolumn{2}{c}{inner} & & \multicolumn{2}{c}{outer}\\
\cline{4-5} \cline{7-8}\\
\colhead{Galaxy}& \colhead{Fit} & \colhead{Filter\tablenotemark{b}} &
\colhead{$\mu_{eff}/\mu_0$\tablenotemark{c}} &\colhead{$R_{eff}/\alpha$\tablenotemark{d}} &
& \colhead{$\mu_0$\tablenotemark{e}} &\colhead{$\alpha$\tablenotemark{f}} &
\colhead{Boundary\tablenotemark{g}} & \colhead{Fit\tablenotemark{h}}
\\
& \colhead{Type \tablenotemark{a}} &
& \colhead{$\left[mag\;arcsec^{-2}\right]$} &\colhead{$\left[^{\prime\prime}\right]$}&
& \colhead{$\left[mag\;arcsec^{-2}\right]$} &\colhead{$\left[^{\prime\prime}\right]$}&
\colhead{$\left[^{\prime\prime}\right]$}&  \colhead{Error}
}
\startdata
\tablenotemark{*}UGC 00023 &    Two Disk &B  & 19.06 (0.33)& 1.71 (0.02) && 21.17 (0.21)&11.05 (0.06)&  6.55  & 0.82 \\
\tablenotemark{*}UGC 00023 &    Two Disk &R  & 17.30 (0.27)& 1.72 (0.02) && 19.58 (0.16)& 9.58 (0.03)&  6.93  & 0.53 \\
   UGC 00023 &  Bulge/Disk &B  & 21.48 (1.98)& 4.06 (0.23) && 21.47 (0.51)&11.96 (0.11)&  8.94  & 0.89 \\
   UGC 00023 &  Bulge/Disk &R  & 19.20 (1.56)& 3.07 (0.14) && 19.82 (0.35)& 9.96 (0.05)&  8.79  & 0.58 \\
\\
\tablenotemark{*}UGC 00189 &    Two Disk &B  & 20.99 (0.08)& 7.75 (0.06) && 24.05 (1.85)&36.90 (2.11)& 28.34  & 0.33 \\
\tablenotemark{*}UGC 00189 &    Two Disk &R  & 19.53 (0.09)& 6.13 (0.05) && 21.66 (0.67)&18.84 (0.22)& 24.92  & 0.97 \\
\\
\tablenotemark{*}UGC 01362 &    One Disk &B  &     \nodata &     \nodata && 23.06 (0.33)& 7.77 (0.10)& \nodata& 1.78 \\
\tablenotemark{*}UGC 01362 &    One Disk &R  &     \nodata &     \nodata && 21.84 (0.15)& 7.85 (0.04)& \nodata& 0.46 \\
\\

\\
\tablenotemark{*}UGC 02299 &    Two Disk &B  & 20.85 (0.10)& 2.00 (0.16) && 22.43 (1.76)&11.24 (0.04)& 6.48   & 1.88 \\
\tablenotemark{*}UGC 02299 &    Two Disk &R  & 19.84 (0.28)& 2.21 (0.03) && 21.46 (0.38)&10.50 (0.10)& 7.23   & 0.99 \\
   UGC 02299 &  Bulge/Disk &B  & 25.10 (2.85)&17.35 (1.95) && 23.23 (2.18)&12.62 (0.27)&17.01   & 2.08 \\
   UGC 02299 &  Bulge/Disk &R  & 24.00 (1.75)&17.76 (1.24) && 22.30 (1.43)&10.37 (0.19)& 0.00   & 1.18 \\
\\
   UGC 02588 &    One Disk &B  &     \nodata &    \nodata  && 21.50 (0.08)& 5.47 (0.02)& \nodata& 1.40 \\
   UGC 02588 &    One Disk &R  &     \nodata &    \nodata  && 20.41 (0.07)& 5.79 (0.02)& \nodata& 1.12 \\
\tablenotemark{*}UGC 02588 &  Bulge/Disk &B  & 25.75 (2.43)&16.16 (1.75) && 22.61 (0.97)& 8.78 (0.08)&  6.94  & 1.18 \\
\tablenotemark{*}UGC 02588 &  Bulge/Disk &R  & 23.77 (2.50)& 7.86 (0.75) && 21.09 (0.48)& 7.41 (0.04)&  4.57  & 0.82 \\

\\
\tablenotemark{*}UGC 02796 &    Two Disk &B  & 19.07 (0.34)& 1.89 (0.03) && 21.02 (0.60)& 8.75 (0.14)&  6.90  & 0.48 \\
\tablenotemark{*}UGC 02796 &    Two Disk &R  & 17.52 (0.27)& 2.19 (0.03) && 19.66 (0.41)& 8.69 (0.08)&  8.59  & 0.68 \\
   UGC 02796 &  Bulge/Disk &B  & 23.07 (2.01)&13.80 (1.10) && 22.54 (4.11)& 8.92 (0.45)& \tablenotemark{\dag} & 0.71 \\
   UGC 02796 &  Bulge/Disk &R  & 21.01 (0.90)&10.26 (0.31) && 20.49 (1.06)& 6.76 (0.22)& \tablenotemark{\dag} & 0.21 \\
\\
   UGC 03119 &    One Disk &B  &     \nodata &    \nodata  && 19.82 (0.07)&10.35 (0.03)& \nodata& 0.75 \\
   UGC 03119 &    One Disk &R  &     \nodata &    \nodata  && 17.92 (0.05)& 9.48 (0.01)& \nodata& 0.49 \\
\tablenotemark{*}UGC 03119 &  Bulge/Disk &B  & 22.05 (9.17)& 3.54 (0.91) && 20.03 (0.31)&11.51 (0.08)&  3.50  & 0.61 \\
\tablenotemark{*}UGC 03119 &  Bulge/Disk &R  & 21.07 (5.81)& 6.53 (1.29) && 18.16 (0.24)&10.10 (0.02)&  3.71  & 0.28 \\
\\
\tablenotemark{*}UGC 03308 &    Two Disk &B  & 19.83 (2.25)& 0.87 (0.04) && 21.70 (0.18)&15.47 (0.16)&  3.47  & 0.35 \\
\tablenotemark{*}UGC 03308 &    Two Disk &R  & 19.06 (1.32)& 0.92 (0.03) && 20.99 (0.13)&11.60 (0.06)&  3.58  & 0.25 \\
   UGC 03308 &  Bulge/Disk &B  & 18.36 (10.1)& 0.43 (0.08) && 21.71 (0.23)&15.66 (0.18)&  3.62  & 0.37 \\
   UGC 03308 &  Bulge/Disk &R  & 18.70 (5.77)& 0.67 (0.08) && 21.03 (0.18)&11.90 (0.08)&  3.85  & 0.20 \\
\\
   UGC 07598 &    One Disk &B  &     \nodata &     \nodata && 21.69 (0.07)& 7.96 (0.02)& \nodata& 0.75 \\
   UGC 07598 &    One Disk &R  &     \nodata &     \nodata && 19.82 (0.08)& 6.26 (0.02)& \nodata& 1.12 \\
\tablenotemark{*}UGC 07598 &  Bulge/Disk &B  & 16.02 (6.98)& 0.21 (0.03) && 21.84 (0.13)& 8.60 (0.03)&  3.25  & 0.52 \\
\tablenotemark{*}UGC 07598 &  Bulge/Disk &R  & 15.78 (3.90)& 0.40 (0.03) && 20.18 (0.21)& 7.34 (0.04)&  4.06  & 0.49 \\
\\
\tablenotemark{*}UGC 08311 &    Two Disk &B  & 20.82 (0.09)& 2.94 (0.02) && 23.29 (0.66)&14.91 (0.26)& 12.50  & 0.48 \\
\tablenotemark{*}UGC 08311 &    Two Disk &R  & 20.28 (0.08)& 3.22 (0.03) && 22.73 (0.71)&15.81 (0.32)& 13.64  & 0.39 \\
\\
\tablenotemark{*}UGC 08644 &    One Disk &B  &     \nodata &    \nodata  && 22.56 (0.14)& 8.79 (0.06)& \nodata& 0.55 \\
\tablenotemark{*}UGC 08644 &    One Disk &R  &     \nodata &    \nodata  && 21.19 (0.15)& 6.40 (0.04)& \nodata& 0.78 \\
\\
\tablenotemark{*}UGC 08904 &    Two Disk &B  & 20.43 (0.16)& 2.56 (0.02) && 22.94 (0.71)&10.07 (0.15)& 11.22  & 0.69 \\
\tablenotemark{*}UGC 08904 &    Two Disk &R  & 18.90 (0.16)& 2.20 (0.01) && 22.05 (0.49)& 9.67 (0.10)& 11.19  & 0.35 \\
   UGC 08904 &  Bulge Only &B  & 23.99 (0.27)&12.55 (0.09) &&     \nodata &    \nodata & \nodata& 1.17 \\
   UGC 08904 &  Bulge Only &R  & 21.83 (0.25)& 6.95 (0.04) &&     \nodata &    \nodata & \nodata& 0.48 \\
\\
\tablenotemark{*}UGC 10894 &    One Disk &B  &     \nodata &    \nodata  && 21.65 (0.10)& 7.90 (0.04)& \nodata& 0.44 \\
\tablenotemark{*}UGC 10894 &    One Disk &R  &     \nodata &    \nodata  && 20.26 (0.12)& 6.69 (0.04)& \nodata& 0.42 \\
\\
   UGC 11068 &    Two Disk &B  & 20.15 (0.65)& 1.15 (0.02) && 22.46 (0.13)&13.65 (0.07)&  4.87  & 0.31 \\
   UGC 11068 &    Two Disk &R  & 18.67 (0.54)& 1.25 (0.02) && 21.02 (0.14)&11.26 (0.05)&  5.23  & 0.54 \\
\tablenotemark{*}UGC 11068 &  Bulge/Disk &B  & 20.36 (0.18)& 1.00 (0.00) && 22.53 (0.11)&14.10 (0.06)&  5.31  & 0.38 \\
\tablenotemark{*}UGC 11068 &  Bulge/Disk &R  & 19.89 (2.52)& 1.66 (0.10) && 21.18 (0.25)&11.97 (0.07)&  6.23  & 0.42 \\
\\
\tablenotemark{*}UGC 11355 &    Two Disk &B  & 19.20 (0.15)& 3.34 (0.02) && 21.61 (0.18)&32.30 (0.24)& 14.23  & 0.86 \\
\tablenotemark{*}UGC 11355 &    Two Disk &R  & 17.30 (0.16)& 3.01 (0.02) && 19.83 (0.13)&24.07 (0.10)& 13.02  & 0.64 \\
   UGC 11355 &  Bulge/Disk &B  & 23.29 (0.95)&19.99 (0.72) && 22.13 (0.56)&38.81 (0.57)& 26.27  & 1.13 \\
   UGC 11355 &  Bulge/Disk &R  & 20.70 (0.94)&10.96 (0.37) && 20.19 (0.35)&26.30 (0.18)& 26.30  & 0.93 \\
\\
\tablenotemark{*}UGC 11396 &    Two Disk &B  & 20.34 (1.82)& 1.14 (0.05) && 22.08 (0.27)&22.66 (0.39)&  4.46  & 0.95 \\
\tablenotemark{*}UGC 11396 &    Two Disk &R  & 19.77 (2.17)& 1.23 (0.08) && 20.20 (0.17)&13.04 (0.10)&  2.80  & 1.69 \\
   UGC 11396 &  Bulge/Disk &B  & 20.52 (9.40)& 0.97 (0.19) && 22.10 (0.35)&23.03 (0.45)&  4.75  & 0.93 \\
   UGC 11396 &  Bulge/Disk &R  & 20.09 (3.23)& 1.09 (0.32) && 20.21 (0.23)&13.03 (0.11)&  3.03  & 1.69 \\
\\
\tablenotemark{*}UGC 11617 &    One Disk &B  &     \nodata &    \nodata  && 21.30 (0.07)&12.83 (0.05)& \nodata& 0.69 \\
\tablenotemark{*}UGC 11617 &    One Disk &R  &     \nodata &    \nodata  && 20.08 (0.08)&10.99 (0.04)& \nodata& 0.46 \\
\\
\tablenotemark{*}UGC 11840 &    No Fit   &B  &     \nodata &    \nodata  &&     \nodata &    \nodata & \nodata& \nodata \\
\tablenotemark{*}UGC 11840 &    No Fit   &R  &     \nodata &    \nodata  &&     \nodata &    \nodata & \nodata& \nodata \\
\\
   UGC 12021 &    One Disk &B  &     \nodata &    \nodata  && 20.73 (0.06)&11.05 (0.02)& \nodata& 1.63 \\
   UGC 12021 &    One Disk &R  &     \nodata &    \nodata  && 19.41 (0.06)&10.26 (0.02)& \nodata& 1.58 \\
\tablenotemark{*}UGC 12021 &  Bulge/Disk &B  & 26.49 (1.28)&23.19 (0.00) && 20.88 (0.16)&11.59 (0.04)&  2.37  & 1.54 \\
\tablenotemark{*}UGC 12021 &  Bulge/Disk &R  & 23.85 (0.32)&23.19 (0.00) && 19.96 (0.18)&11.60 (0.03)&  6.23  & 1.01 \\
\\
\\

\enddata
\tablecomments{Derivation of quantities is described in Section~\ref{sec:reduce}. Errors are given in parenthesis.}
\tablenotetext{a}{The type of fit made to the surface brightenss profile -- bulge+disk, two exponential disks, or one exponential disk.}
\tablenotetext{b}{Optical filter for the data described within that row.}
\tablenotetext{c}{Effective surface brightness (R$^{1/4}$ bulge fit) or central surface brightness (exponential disk fit) for the inner disk fit. See Equations~\ref{eqn:bulge} and \ref{eqn:disk}. }
\tablenotetext{d}{Effective radius (R$^{1/4}$ bulge fit) or scale length (exponential disk fit) for the inner disk fit.
See Equations~\ref{eqn:bulge} and \ref{eqn:disk}.}
\tablenotetext{e}{Central surface brightness for the outer exponential disk fit.}
\tablenotetext{f}{Scale length for the outer exponential disk fit.}
\tablenotetext{g}{Boundary between the inner and outer fits, defined by where the fitted lines cross.}
\tablenotetext{h}{$\chi^2$ error for the fits.}
\tablenotetext{*}{Best fit -- used for all further analysis.}
\tablenotetext{\dag}{Fitted lines for the bulge and disk components do not cross.}
\end{deluxetable}

%
%
\begin{deluxetable}{cccccccccc}
\rotate
\tabletypesize{\scriptsize}
\tablewidth{0pt}
\tablecaption{Properties of H$\alpha$ Regions\label{tab:regions}}
\tablehead{
\colhead{Galaxy}& \colhead{Region\tablenotemark{a}} &\colhead{\Ha\ Flux
$\times 10^{-15}$} &
\colhead{\Ha\ Luminosity$\times 10^{38}$} & \colhead{SFR\tablenotemark{b}} &
\colhead{EW\tablenotemark{c}} & \colhead{B\tablenotemark{d}}
&\colhead{R\tablenotemark{d}} &
\colhead{B$-$R\tablenotemark{d}} &\colhead{Diffuse\tablenotemark{e}}
\\
&&\colhead{$\left[erg\; cm^{-2}\; s^{-1}\right]$} & \colhead{$\left[erg\;
s^{-1}\right]$}&
\colhead{$\left[M_\odot\; yr^{-1}\right]$} & \colhead{$\rm \left[\AA \right]$}
&
\colhead{$\left[mag\right]$} & \colhead{$\left[mag\right]$} &
\colhead{$\left[mag\right]$} & \colhead{$\left[\%\right]$}
}
\startdata

UGC 00023 &	1 &	70	(14)&	760	(150)&	0.6	(0.1)&	23	(1)&	16.9	(0.1)&	15.2	(0.1)&	1.7	(0.2)&	0.82	(0.04) \\
UGC 00189 &	1 &	10	(2)&	110	(20)&	0.09	(0.02)&	30	(2)&	19.9	(0.6)&	18.3	(0.5)&	1.7	(0.7)&	0.98	(0.01) \\
UGC 01362 &	0 &	\nodata	&	\nodata	&	\nodata &	\nodata	&	\nodata	&	\nodata	&	\nodata	&	\nodata \\
UGC 02299 &	1 &	12	(2)&	240	(50)&	0.19	(0.04)&	20	(1)&	21	(2)&	21	(2)&	0.	(3)&	\nodata \\
UGC 02299 &	2 &	15	(3)&	260	(50)&	0.21	(0.04)&	14	(1)&	18.2	(0.2)&	17.1	(0.2)&	1.1	(0.3)&	\nodata \\
UGC 02299 &	TOTAL &	22	(3)&	490	(70)&	0.39	(0.06)&	\nodata &	\nodata &	\nodata &	\nodata &	0.45	(0.06) \\
UGC 02588 &	0 &	\nodata	&	\nodata	&	\nodata &	\nodata	&	\nodata	&	\nodata	&	\nodata	&	\nodata \\
UGC 02796 &	1 &	74.8	(15)&	900	(200)&	0.71	(0.1)&	10	(1)&	16.5	(0.2)&	14.8	(0.4)&	1.7	(0.4)&	0.67	(0.01) \\
UGC 03119 &	1 &	151	(30)&	1600	(300)&	1.3	(0.3)&	16	(1)&	18.8	(0.1)&	16.7	(0.1)&	2	(0.1)&	0.77	(0.02) \\
UGC 03308 &	0 &	\nodata	&	\nodata	&	\nodata &	\nodata	&	\nodata	&	\nodata	&	\nodata	&	\nodata \\
UGC 07598 &	1 &	6	(1)&	80	(17)&	0.07	(0.01)&	13	(1)&	22.2	(0.7)&	21.5	(0.7)&	0.6	(1)&	\nodata \\
UGC 07598 &	2 &	4.7	(0.9)&	60	(12)&	0.05	(0.01)&	12	(1)&	22	(1)&	20.2	(0.9)&	2	(1)&	\nodata \\
UGC 07598 &	3 &	2.4	(0.5)&	32	(6)&	0.03	(0.01)&	13	(1)&	23	(1)&	21	(1)&	2	(2)&	\nodata \\
UGC 07598 &	4 &	6	(1)&	80	(20)&	0.06	(0.01)&	12	(1)&	21.1	(0.8)&	19.6	(0.7)&	2	(1)&	\nodata \\
UGC 07598 &	5 &	3.3	(0.7)&	43	(9)&	0.03	(0.01)&	13	(1)&	23	(2)&	22	(2)&	2	(3)&	\nodata \\
UGC 07598 &	6 &	6	(1)&	80	(20)&	0.06	(0.01)&	11	(1)&	21	(3)&	22	(3)&	-1	(4)&	\nodata \\
UGC 07598 &	7 &	4.1	(0.8)&	50	(10)&	0.04	(0.01)&	11	(1)&	22	(1)&	21	(2)&	1	(4)&	\nodata \\
UGC 07598 &	8 &	1.3	(0.3)&	15	(3)&	0.012	(0.003)&	9	(1)&	25	(3)&	24	(3)&	1	(3)&	\nodata \\
UGC 07598 &	9 &	32	(6)&	430	(90)&	0.34	(0.07)&	13	(1)&	18	(0.1)&	16	(0.1)&	1.9	(3)&	\nodata \\
UGC 07598 &	10 &	1.4	(0.3)&	20.	(4)&	0.015	(0.003)&	16	(1)&	21.8	(0.5)&	20.4	(0.5)&	1.3	(0.5)&	\nodata \\
UGC 07598 &	11 &	10	(2)&	130	(30)&	0.11	(0.02)&	11	(1)&	21.4	(0.5)&	20	(0.6)&	1.4	(0.8)&	\nodata \\
UGC 07598 &	12 &	4.8	(1)&	60	(10)&	0.05	(0.01)&	11	(1)&	22	(1)&	20.2	(0.9)&	2	(1)&	\nodata \\
UGC 07598 &	13 &	4.4	(0.9)&	60	(10)&	0.05	(0.01)&	12	(1)&	21.8	(0.8)&	20.1	(0.7)&	2	(1)&	\nodata \\
UGC 07598 &	14 &	5	(1)&	70	(10)&	0.05	(0.01)&	11	(1)&	22.7	(1)&	21	(1)&	1	(2)&	\nodata \\
UGC 07598 &	TOTAL &	69	(6)&	1200	(100)&	0.96	(0.08)&	\nodata &	\nodata &	\nodata &	\nodata &	0.5	(0.1) \\
UGC 08311 &	1 &	7	(1)&	14	(3)&	0.011	(0.002)&	20.	(1)&	21.3	(0.7)&	21	(0.9)&	0.	(1)&	\nodata \\
UGC 08311 &	2 &	100	(20)&	240	(50)&	0.19	(0.04)&	53	(1)&	18.8	(0.4)&	18.4	(0.5)&	0.3	(0.6)&	\nodata \\
UGC 08311 &	3 &	7	(1)&	16	(3)&	0.012	(0.003)&	18	(1)&	20.8	(0.1)&	20.	(1)&	1	(1)&	\nodata \\
UGC 08311 &	4 &	17	(3)&	38	(8)&	0.03	(0.01)&	29	(1)&	0.9	(0.1)&	20.5	(0.9)&	-19.6	(0.9)&	\nodata \\
UGC 08311 &	5 &	8	(2)&	18	(4)&	0.015	(0.003)&	25	(1)&	22	(2)&	22	(2)&	1	(3)&	\nodata \\
UGC 08311 &	6 &	140	(30)&	320	(60)&	0.26	(0.05)&	38	(1)&	16.9	(0.1)&	16.3	(0.1)&	0.6	(0.1)&	\nodata \\
UGC 08311 &	7 &	15	(3)&	32	(6)&	0.03	(0.01)&	19	(1)&	18.6	(0.1)&	18.1	(0.2)&	0.6	(0.2)&	\nodata \\
UGC 08311 &	TOTAL &	270	(30)&	680	(80)&	0.54	(0.07)&	\nodata &	\nodata &	\nodata &	\nodata &	0.2	(0.1) \\
UGC 08644 &	1 &	5.1	(1)&	44	(9)&	0.04	(0.01)&	17	(1)&	21	(0.1)&	20.	(1)&	1	(1)&	\nodata \\
UGC 08644 &	2 &	4.8	(1)&	42	(8)&	0.03	(0.01)&	20.	(1)&	22	(0.5)&	21.1	(0.6)&	0.9	(0.8)&	\nodata \\
UGC 08644 &	TOTAL &	8.3	(1)&	90	(10)&	0.07	(0.01)&	\nodata &	\nodata &	\nodata &	\nodata &	\nodata \\
UGC 08904 &	1 &	3.4	(0.7)&	47	(9)&	0.04	(0.01)&	10.	(1)&	22.4	(0.8)&	22	(1)&	1	(1)&	\nodata \\
UGC 08904 &	2 &	0.7	(0.1)&	12	(2)&	0.01	(0.009)&	23	(1)&	23.6	(0.5)&	23.0	(0.6)&	0.7	(0.8)&	\nodata \\
UGC 08904 &	3 &	1.2	(0.3)&	20.	(4)&	0.02	(0.02)&	16	(1)&	24.3	(0.7)&	24	(1)&	1	(1)&	\nodata \\
UGC 08904 &	4 &	0.8	(0.2)&	12	(2)&	0.01	(0.009)&	10.	(1)&	23.8	(0.4)&	24	(1)&	0.	(1)&	\nodata \\
UGC 08904 &	5 &	0.7	(0.1)&	9	(2)&	0.01	(0.007)&	10.	(1)&	24.5	(0.9)&	24	(1)&	1	(2)&	\nodata \\
UGC 08904 &	6 &	32	(6)&	600	(100)&	0.45	(0.09)&	25	(1)&	17.7	(0.1)&	16.3	(0.1)&	1.4	(0.1)&	\nodata \\
UGC 08904 &	TOTAL &	33	(6)&	700	(100)&	0.53	(0.09)&	\nodata &	\nodata &	\nodata &	\nodata &	0.6	(0.1) \\
UGC 10894 &	1 &	1.2	(0.2)&	11	(2)&	0.009	(0.002)&	39	(3)&	24	(2)&	23	(1)&	1	(2)&	\nodata \\
UGC 10894 &	2 &	1.4	(0.3)&	13	(3)&	0.010	(0.002)&	37	(2)&	22.1	(0.5)&	21.3	(0.6)&	0.8	(0.8)&	\nodata \\
UGC 10894 &	3 &	13	(3)&	110	(20)&	0.09	(0.02)&	27	(1)&	18.9	(0.2)&	17.2	(0.1)&	1.8	(0.2)&	\nodata \\
UGC 10894 &	4 &	1.9	(0.4)&	18	(4)&	0.014	(0.003)&	31	(2)&	23	(1)&	22	(1)&	1	(2)&	\nodata \\
UGC 10894 &	5 &	1.8	(0.4)&	16	(3)&	0.013	(0.003)&	35	(2)&	22.8	(0.7)&	21.9	(0.7)&	1	(1)&	\nodata \\
UGC 10894 &	6 &	2.8	(0.6)&	26	(5)&	0.020	(0.004)&	36	(2)&	22.4	(0.7)&	22	(1)&	0.	(1)&	\nodata \\
UGC 10894 &	7 &	3.9	(0.8)&	35	(7)&	0.03	(0.01)&	31	(2)&	22.6	(0.8)&	21.6	(0.8)&	1	(1)&	\nodata \\
UGC 10894 &	TOTAL &	23	(2)&	230	(25)&	0.18	(0.02)&	\nodata &	\nodata &	\nodata &	\nodata &	0.6	(0.3) \\
UGC 11068 &	0 &	\nodata	&	\nodata	&	\nodata &	\nodata	&	\nodata	&	\nodata	&	\nodata	&	\nodata \\
UGC 11355 &	1 &	180	(40)&	600	(100)&	0.48	(0.1)&	17	(1)&	19.7	(0.4)&	18.2	(0.4)&	1.5	(0.6)&	0.87	(0.08) \\
UGC 11396 &	1 &	25	(5)&	90	(20)&	0.07	(0.01)&	21	(1)&	22	(1)&	18.1	(0.2)&	4	(1)&	0.91	(0.01) \\
UGC 11617 &	1 &	9	(2)&	40	(8)&	0.07	(0.01)&	12	(1)&	21	(1)&	21	(0.6)&	0.	(1)&	\nodata \\
UGC 11617 &	2 &	21	(4)&	90	(20)&	0.04	(0.01)&	13	(1)&	21	(2)&	20.	(2)&	1	(3)&	\nodata \\
UGC 11617 &	3 &	13	(3)&	50	(10)&	0.04	(0.01)&	13	(1)&	22	(2)&	20.	(3)&	1	(3)&	\nodata \\
UGC 11617 &	4 &	12	(2)&	47	(9)&	0.09	(0.01)&	11	(1)&	21	(2)&	21	(1)&	1	(3)&	\nodata \\
UGC 11617 &	5 &	26	(5)&	110	(20)&	0.03	(0.02)&	14	(1)&	21.5	(0.6)&	21	(4)&	1	(4)&	\nodata \\
UGC 11617 &	6 &	10	(2)&	42	(8)&	0.03	(0.01)&	12	(1)&	21.5	(0.2)&	20.8	(0.7)&	0.7	(0.7)&	\nodata \\
UGC 11617 &	TOTAL &	69	(6)&	390	(30)&	0.31	(0.03)&	\nodata &	\nodata &	\nodata &	\nodata &	0.8	(0.1) \\
UGC 11840 &	0 &	\nodata	&	\nodata	&	\nodata &	\nodata	&	\nodata	&	\nodata	&	\nodata	&	\nodata \\
UGC 12021 &	1 &	13	(3)&	50	(10)&	0.04	(0.01)&	31	(2)&	19.2	(0.4)&	17.8	(0.3)&	1.4	(0.5)&	0.97	(0.01) \\
\enddata
\tablecomments{Derivation of quantities is described in
Section~\ref{sec:reduce}. Errors are given in parenthesis.}
\tablenotetext{a}{Internal numbering scheme for the HII regions found by
HIIphot.}
\tablenotetext{b}{SFR for the region, defined as SFR = ${ {L_{H\alpha}} \over
{1.26 \times 10^{41} erg\;s^{-1} } }$.}
\tablenotetext{c}{Equivalent width was calculated simply as the ratio of the
total \Ha\ flux to total \Ha-subtracted continuum flux.}
\tablenotetext{d}{Total B and R magnitudes and colors within the HII regions}
\tablenotetext{e}{Diffuse fraction found for the galaxy, defined defined
here as the ratio of the fraction of \Ha\ flux not found within the defined
\Ha\ regions to the total \Ha\
flux found for the entire galaxy.}
\tablenotetext{\dag}{Due to both a (masked) star near the center of this
galaxy and a (masked) CCD flaw
(bad column) which also runs through the center of the galaxy, a number of
\ion{H}{2} regions which
should have been identified by HIIphot were not, artificially rasing the
diffuse fraction on this galaxy,
possibly by as much as 20-30\%.
}
\end{deluxetable}

\clearpage

%
%
%
%
%
%

\begin{figure}
\plotone{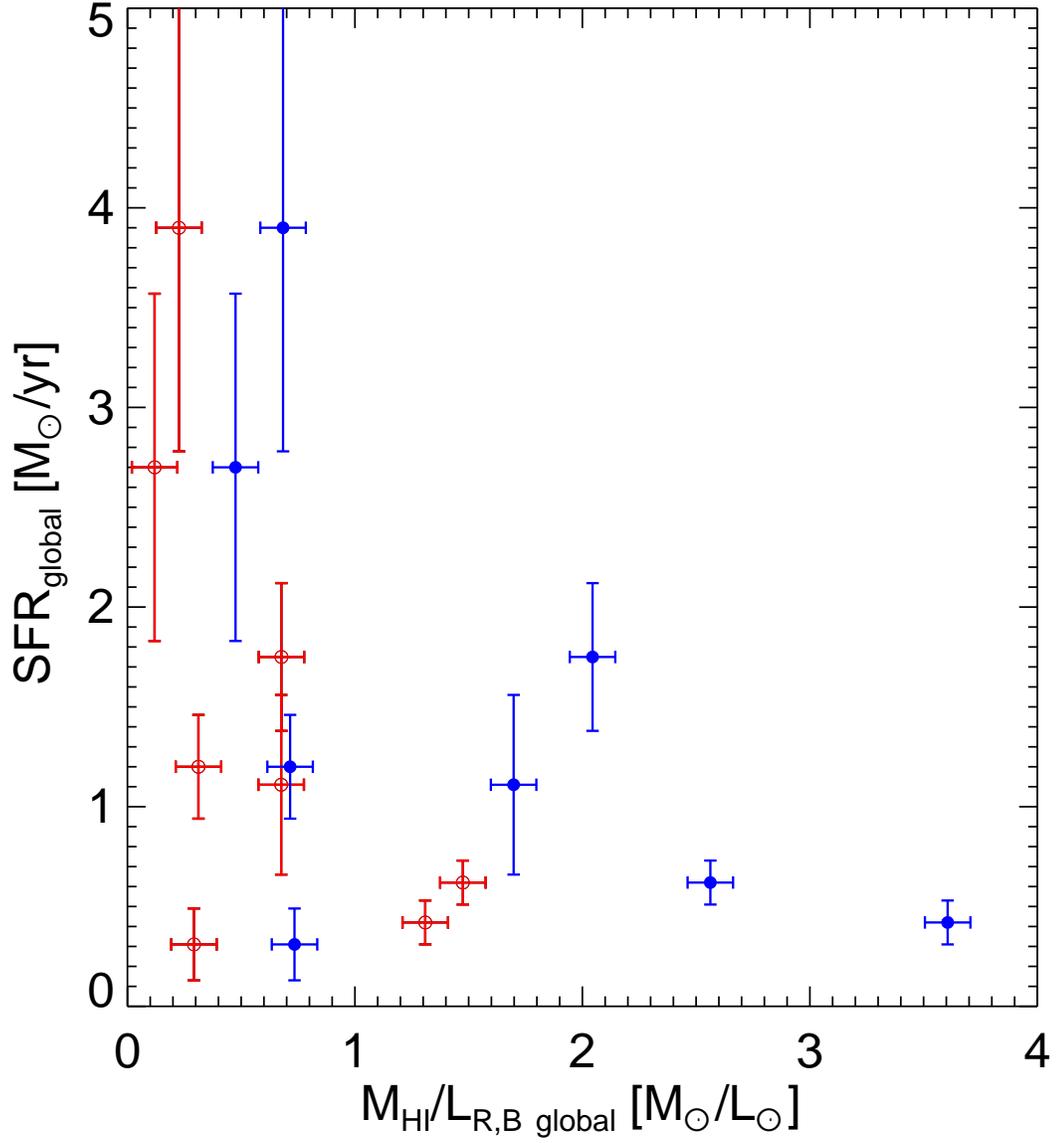}
\caption{Gas mass to B and R-band luminosity ratios plotted against the global star formation rate for the galaxies.
The (blue) filled circles are for the B-band data and the (red) open circles are for the R-band data. \label{fig:MHIL_SFR} }
\end{figure}

\end{document}